\documentclass{jpp}
\usepackage{graphicx}
\usepackage{epstopdf, epsfig}
\usepackage{hyperref}

\usepackage{graphicx,natbib,color}
\usepackage{amssymb}
\usepackage[T1]{fontenc}
\usepackage[latin1]{inputenc}
\usepackage{color}

\usepackage{enumitem}

\def\be{\begin{equation}}
\def\ee{\end{equation}}
\def\bea{\begin{eqnarray}}
\newcommand{\cc}{\mathrm{C}}

\newcommand{\apj}{Astrophys.\ J.\ }
\newcommand{\mnras}{MNRAS}

\newcommand{\prd}{Phys.\ Rev.\ \textbf{D}}

\newcommand{\aap}{A\&A}
\newcommand{\Qa}{\mathcal{Q}}
\newcommand{\Sa}{\mathcal{S}}
\newcommand{\la}{\mathcal{A}}

\def\eea{\end{eqnarray}}

\def\Sie{\mathcal{S}}

\newcommand{\pp}{\textbf{()}}

\newcommand{\mso}{\mathrm{mso}}

\newcommand{\il}{~}
\newcommand{\ti}[1]{\mbox{\tiny{#1}}}
\newcommand{\g}[1]{\Gamma^{\phantom\ #1}}
\newcommand{\rc}{\rho_{\ti{C}}}
\newcommand{\Mie}{\mathcal{M}}
\newcommand{\dd}{\mathrm{d}}

\shorttitle{Aspects of GR-MHD in High-Energy Astrophysics}
\shortauthor{D. Pugliese and G. Montani}

\title{Aspects of GR-MHD in High-Energy Astrophysics}

\author{Daniela Pugliese\aff{1}
  \corresp{\email{d.pugliese.physics@gmail.com}},
 \and Giovanni Montani\aff{2}}

\affiliation{\aff{1}Research Centre of Theoretical Physics and Astrophysics,
Institute of Physics,
  Silesian University in Opava,
 Bezru\v{c}ovo n\'{a}m\v{e}st\'{i} 13, CZ-74601 Opava, Czech Republic
\aff{2}ENEA- R.C. Frascati, UTFUS-MAG, Via Enrico Fermi 45, Frascati, Roma 00044, Italy\\
Physics Department, "Sapienza" University of Rome, P.le Aldo Moro 5, Roma 00185, Italy}

\begin{document}

\maketitle

\begin{abstract}
 This  work focuses on some key aspects
   of  the General Relativistic (GR) - magneto-hydrodynamic (MHD) applications  in High-Energy Astrophysics.
 We discuss the  relevance of   the  GR-HD  counterparts   formulation
 exploring the  geometrically thick disk models and  constraints of the
GR-MHD  shaping  the  physics of accreting configurations.
Models of  clusters of   tori orbiting a central super-massive black hole (\textbf{SMBH}) are described.   These  orbiting tori aggregates form   sets of  geometrically thick, pressure supported,  perfect fluid tori, associated to  complex  instability processes including  tori collision  emergence and empowering a wide range of activities related expectantly to the embedding matter environment of   Active Galaxy Nuclei.
Some notes are  included
on  aggregates  combined with   proto-jets, represented by open cusped solutions  associated to the geometrically thick   tori.
This exploration of some key concepts of the GR-MHD  formulation in its applications to High-Energy Astrophysics starts with the discussion of   the initial data problem for a most general Einstein-Euler-Maxwell system  addressing  the problem with a relativistic geometric background. The system is then set in  quasi linear  hyperbolic form, and the reduction procedure is argumented.
Then, considerations follow on the analysis of the stability problem for self-gravitating systems with determined symmetries considering the perturbations also of the  geometry part  on the quasi  linear hyperbolic onset.
Thus we focus on the ideal  GR-MHD and self-gravitating  plasma ball.
We conclude with the  models of geometrically thick GR-HD disks gravitating around a  Kerr \textbf{SMBH} in their GR-HD formulation and  including   in the force balance equation of the disks  the influence of a toroidal magnetic field,  determining  its impact in  tori topology and stability.
\end{abstract}

\section{Introduction}
High-Energy Astrophysics  is characterized by   several   processes  involving different attractors as  stars  and super-massive black holes (\textbf{SMBHs}).
These strong attractors,  among the most energetic objects in the Universe, are supposed to be the  engines  of the high-energy outbursts  of matter and fields, including   $\gamma$-rays  and  X-rays  emissions. The  modelizations  of the related   observations and  the description of the engines empowering these processes are  dealt with a great  computer simulation effort and differently challenging  analytical means.
These studies also  involve very complex issues related to the life of the  attractor as   the \textbf{SMBH} formations and the progenitor collapse towards the  formations  of the \textbf{BH} horizons,  the energy extraction  from \textbf{BHs} and \textbf{BHs} evolutions,  the  exact mechanism empowering accretion  and jet emission.
Even the identification of the  \textbf{BH} through its  (ADM)  mass and spin is a debated issue.
Many methods are  connected to the  accretion disk physics,  dealing with  luminosity, accretion rates, location of inner edge and energy extraction. All these   approaches are in large  extents  model dependent, as the    same definitions of key aspects of \textbf{BHs} accretion, as the  disks inner edge, are  not at all clearly  established.
Recent and fairly independent methods includes  the Gravitational Wave detection.
 Others, still rather mysterious phenomena, related to \textbf{BHs} accretion are  the   high-frequency quasi periodic oscillations (\textbf{QPOs})
observed in non-thermal X-ray emission from compact
objects.

Active Galactic Nucleai (\textbf{AGNs}) provide a rich scenario to observe \textbf{SMBHs} interacting with
their environments.
These embeddings  can host the  formation  of \textbf{BH} accretion disks  with a structured  inner  morphology,  following  evolution of the  central \textbf{BH}  attractor interacting with the surrounding masses during several
accretion regimes occurred in the lifetime of non-isolated Kerr \textbf{BHs}.
The inner structure of  the orbiting disk may leave traces of chaotical, discontinuous accretion episodes
 in the form of matter remnants orbiting the
central attractor producing sequences of orbiting toroids
 with strongly different features as different rotation
orientations with respect to the Kerr \textbf{BH}. Consequently   corotating and
counterrotating accretion stages can be mixed.
This situation has been  modelled in  Ringed accretion disks (\textbf{RAD}) in  \cite{pugtot,ringed,open,dsystem,long,Multy,proto-jet,letter,Fi-Ringed}.
Ringed disks are  clusters of  toroidal  (thick disk) configurations  centered on a single  \textbf{BH}, and prescribed by  barotropic models.
The model strongly
binds the fluid and \textbf{BH} characteristics providing
indications on the situations where to search for
\textbf{RADs} observational evidences. The instability and the \textbf{RAD} structure  strongly  depend on the dimensionless
spin of the  central spinning   attractor. These systems, were  developed in  special   cases as  \textbf{eRAD}, that is \textbf{RADs}  composed by  tori sharing the same equatorial plane which is also their symmetry plane and the equatorial plane of the central attractor. The \textbf{eRAD}  admits  a  number of the instability points related to onset of GR-HD-pressure -supported accretion which is  generally
limited to n=2.
From phenomenological view point these structures have been connected to different processes, for example the obscuring and screening tori, possibly evident as traces
(screening) in X-ray spectrum emission.   More generally however,
a \textbf{RAD} observational evidence is expected by the spectral features of
\textbf{AGNs} X-ray emission shape, due to X-ray obscuration  by one of the tori, providing a \textbf{RAD} fingerprint as a
radially stratified emission profile. This discussion of tori accreting disks is the focus of a part of our investigations.

This work  has been  intended as a conceptual review from the  problematic and  fundamental topics of the GR-MHD (general relativistic magnetohydrodynamic) systems and relevance of their GR-HD  counterparts in dealing with some of the  complex aspects of the Relativistic Astrophysics.
We review    topics  of GR-MHD setup  for a diversified set of scenarios, particularly on regards of the problem of existence and uniqueness of the solutions and their stability --Sec.\il(\ref{Sec:trob}). Therefore we
  look into  the problems  related to the  perturbation of the gravitational field in the self-gravitating systems, the thermodynamic conditions implied by these  in  Sec.\il(\ref{Sec:terms-cons}), and  the propagation of the constraints (addressed in a brief discussion in Sec.\il(\ref{Sec:trob})). The
  constraints  are treated in the context of the accretion disks and particularly in the constrained models of thick disks where the symmetries imply an annulment of the evolution equations --Sec.\il(\ref{Sec:la-stafdisc-gtr})-(\ref{Sec:constr}). We inspect  the issue of   the multiple accreting  systems  in Sec.\il(\ref{Sec:intro-RAD}).
 A goal of this review is therefore  to
 examine  the   validity of the assumptions on matter fields into  constitutive equations and state equations--Sec.\il(\ref{Sec:trob})-Sec.\il(\ref{Sec:prez}), the implications on fluids thermodynamics, Sec.\il(\ref{Sec:terms-cons}), especially in the context of fixed symmetries in self-gravitating objects, and  toroid  symmetry on a fixed background- - Sec.\il(\ref{Sec:influence}).

This brief  review is  conveniently  divided into two parts. We start addressing the  general Einstein--Maxwell--Euler (EME) system.
 In  the first part  the discussion focuses  on some fundamental aspects  of the  MHD  when considered in its full GR formulations and the GR-HD counterparts for the description of accretion toroids around  very compact objects.
Part I, consisting of Sec.\il(\ref{Sec:ideal}), includes a discussion of the   initial data problem for a general EME system.  The reduction procedure is implemented  and the system is  set in  quasi linear  hyperbolic form.
Considerations follow on the analysis of the stability problem for self-gravitating systems with determined symmetries considering the perturbations also of the  geometry part  on the quasi  linear hyperbolic onset.

More in details: in Sec.\il(\ref{Sec:trob}) we discuss  the  initial value problem and stability analysis for the GR-MHD systems, addressing the
formulation of an
initial value problem, the stability analysis for  systems with a privileged direction of symmetry and the gauge conditions
and techniques on stability analysis. (The self-gravitating fluid configuration we selected   can be finalized in a spherical self-gravitating   radially pulsating ball
or characterized in a more complex combination of modes).
Problem set-up and the   equations of the GR-MHD systems are  in Sec.\il(\ref{Sec:prez}) while in Sec.\il(\ref{Sec:terms-cons})
thermodynamic considerations follows. Finally this first part concludes  in Sec.\il(\ref{Subsection:MHD}) with a summary of the situation for  the
infinitely conductive plasma.

The review continues with part II,  Sec.\il(\ref{Sec:Stationar})  and  Sec.\il(\ref{Sec:influence}), in which we  consider orbiting systems on a fixed background with simplifying conditions on the system of equations and symmetries, especially the accretion thick  disks  for very compact objects, where the gravitational part of the disk force balance equation  dictates the geometry and has a predominant role in determining some basic characteristics of the systems.
The  aggregate of tori is modeled as a single   orbiting configuration, by  introducing  a leading function governing the distribution of toroids around  the  black hole attractor. (Eventually, disks agglomerate  composed by tilted tori can be seen, depending on the     tori thickness, and other conditions on tori and a ringed, \textbf{RAD}, structure as  (multipole) gobules of orbiting matter, with different toroidal   spin orientations, covering the  embedded central black hole).
These systems are shown to include   tori with emerging instability phase  related to accretion onto the central black hole and can   provide even   an  evaluation  of  quantities  related  to tori energetics such as the  mass-flux,  the  enthalpy-flux,
and  the flux thickness, depending on the model parameters for tori composed of
polytropic fluids \citep{letter}.

In details: The GR-HD systems of  thick   tori  orbiting  Kerr \textbf{SMBHs}
are introduced in Sec.\il(\ref{Sec:Stationar}) where
 we start introducing  the Kerr geometry in Sec.\il(\ref{Sec:kerr-flji}).
 The HD-"stationary" configurations for thick tori are presented in Sec.\il(\ref{Sec:la-stafdisc-gtr}).
 Constraints, morphology and angular momentum distribution in the disks are discussed in Sec.\il(\ref{Sec:constr}).
 Multi-accretion processes, multi--orbiting structures  and ringed accretion disks are the focus of Sec.\il(\ref{Sec:intro-RAD}).
 This second part of the review  ends in Sec.\il(\ref{Sec:influence}) dealing with
 magnetized tori, considering  toroidal magnetic field in multi-accreting tori.

The study of the GR-HD counterpart in providing constraints for the development of more complex systems  was motivated by the fact that, despite  all the complexity of GR-MRI (magneto-rotational-instability) basing accretion processes in many GR-MHD  disks - many   features and properties are well described globally and  morphologically by the HD models\citep{Shafee,Fragile:2007dk,DeVilliers,Luci,abrafra}.
We have considered  the Polish doughnut (PD) model   of thick disk  for a   perfect fluid, circularly orbiting around a black hole of Schwarzschild and Kerr, in an  hydrodynamic and magnetohydrodynamic context, using the effective potential  approach examined in detail for the  hydrodynamical model in \citet{mnras1} We used the solutions of the barotropic torus with a toroidal magnetic field  discussed in \citet{Komissarov:2006nz,Montero:2007tc}. We have considered the magnetic contribution to the total pressure of the fluid as a perturbation of the effective potential for the exact full general relativistic part.
Relativistic hydrodynamics proved    a successful framework of analysis an frame of contraints for more refined systems.
As  illustrated in a number of diversified scenarios,  the relativistic HD plays a major role also  in the presence of magnetic fields.
\section{On the GR-MHD and the ideal GR-MHD}\label{Sec:ideal}
General Relativistic Magnetohydrodynamic
({GR-MHD}) is a complex set up for the description of a  wide   number of different aspects of High Energy Physics and High Energy  Astrophysics where the GR component of the MHD onset   is relevant in determining  the stability and morphology of the configurations. These systems comprise self-gravitating  objects, where the  GR is a part of the equations to be treated as variables and within the scheme of the perturbation analysis, and  extended matter orbiting   on fixed backgrounds. In this {first part} of the review we   consider the general set-up adapted to  general self-gravitating systems.  Although the analysis of gravitating  systems such as accretion disks are often analyzed  with emerging self-gravity effects,  in  Secs\il(\ref{Sec:Stationar}) and (\ref{Sec:influence}) we focus on the disk gravitating  around a spinning \textbf{SMBH}.
Starting from a self-gravitating onset,  equations  and perturbations can be easily found  in the fixed-background  geometry  case.

We start by considering   systems  described by the coupled EME equations, which include   general relativistic  magnetohydrodynamic (MHD)  systems.  The perturbation  analysis of such systems,  requiring  numerical  approaches,  is often very complicated, even if  often implemented with  suitable assumptions on the  symmetries and the  dynamics.
Generally,  MHD models find  different problematics  when  considered with  gravity. Firstly  there is the  more general  and fundamental issue of  the well-posedness of the equations with regards  to the system initial data problem. Secondly and connected to this first point is the onset of the  stability analysis for systems  with given  symmetries.
We  explore the first point   in Sec.\il(\ref{Sec:trob}).

GR-MHD is involved in a very large number phenomena of  accretion  disks physics,  from the  magnetosphere  configuration to the \textbf{BH} energy extraction.
In  a variety  of astrophysical scenarios the stability problem of plasma configurations is a crucial  issue
 involved in several aspects of High-Energy Astrophysics,  ranging from  stellar objects  to  accretion
disks  and their associated processes, and
 grounding in many models the instabilities describing the accretion processes  or the jet emission.  On the other hand, it is also clear that different models of accretion disk systems and  attractors, imply extremely  different typical  instabilities,  ranging from the MRI (magneto-rotational-instability) processes for example applied to accretion,   to PPI  (Papaloizou-Pringle Instability), an  instabilities  typically of geometrically thick disks  and often interacting with MRI,  to the  runaway instability. We  discuss these aspects further in Sec.\il(\ref{Sec:Stationar}) and Sec.\il(\ref{Sec:influence}).
Due to the complexity of the
 equations, the analysis of the stability and the shape
of accretion disks and flows is often addressed by numerical methods.
A key  challenge  in dealing with   the construction of the  numerical solutions is to reduce the system  to  an appropriate formulation of the well-posedness problem of the  initial values
 and the stability problem. The system   formulation should ensure the   local
and global existence problems \citep{first,ShiSek05,BauSha03,PalGarLehLie10,Anile89,Lichnerowicz67,Disconzi(2014),Radice:2013xpa,Anton2006}.

\medskip

In this section,  and Sec.\il(\ref{Sec:Stationar}) for the analysis of accreting disks,   we consider a barotropic equation of state (\textbf{EoS}).
 As we shall discuss in Sec.\il(\ref{Sec:terms-cons}),
 when the fluid entropy is a constant
of both space and time,  an equation of state to link the pressure $p$
to the matter density $\rho$ can be given in the form $p = p(\rho)$
--- \citep{first}. In this analysis we restrict our
discussion to the case  of isotropic fluids, considering  a one species particles
fluid (sometimes known as {simple fluid}). It is assumed  that  there is  no particle
annihilation or creation leading to the equation of
conservation of particle number.
In parts of this discussion we shall assume a polytropic
equation of state and  a constant velocity of sound.  To
simplify  the  analysis of the stability problem of
linear perturbations, it will be necessary to  specify the
form of the conduction current using the  {Ohm's law}. This assumption ensures that  linear relation between the conduction current
and the electric field holds.  Thus  considering  isotropic
fluids   we can set  a constant electrical
conductivity coefficient for the fluid (plasma).
\subsection{GR-MHD systems: initial value problem and stability analysis}\label{Sec:trob}
\textbf{Formulation of an
initial value problem. }

  The formulation of a good  initial value problem is a
 relevant issue  of  the
  MHD applications,  including  systems described in  their  GR reformulation for self-gravitating and orbiting systems.
 The formulation of an
initial value problem is a necessary starting point for the
construction of numerical solutions of  the
EME system.
General Relativity  admits itself an  initial value problem
formulation which is   its own    Cauchy problem.
  In the GR frame, prescribing   initial data on a
3-dimensional hypersurface, one could  reconstruct the
spacetime associated to this initial data.
Having   a well  initial value problem
formulation  implies some  unquestionable advantages,   it ensures  the  (local and global) existence and uniqueness of the solution, starting from initial data and secondly it allows a  correct and convenient procedure for the   stability analysis of certain reference solutions.
Here we follow the  procedure  in \citet{first} implemented for {GR-MHD} system where  this problem was addressed with the construction of a first order symmetric
 quasi-linear hyperbolic evolution system for the EME system
 (evolution equations) describing
a charged ideal perfect fluid coupled to the Einstein field equations.
The analysis in  \citet{first} shows how the
EME system admits a reduction procedure leading to
symmetric hyperbolic evolution (first-order) equations and thus this automatically shows the  local existence and uniqueness of the   solutions for the
system under consideration.
The analysis we review in this section, is based on a $(1+3)$-tetrad  decomposition,  in order to
ensure the symmetric hyperbolicity of the evolution equations implied
by the Bianchi identity,  and using   tensor of rank 3 corresponding
to the covariant derivative of the Faraday tensor as auxiliary function,  making use of the
Weyl  tensor components  as one of the unknowns.  The system can be further reduced,
assuming the  case of a perfect fluid with infinite conductivity (ideal
MHD) as a particular subcase, which applies
to the study of plasmas in situations where a fluid is
subjected to significant magnetic fields.
This approach serves especially when the gravity  is perturbed.
The evolution equations have unique  local solution (a solution
that exists at least for  a small interval of time),
the solutions obtained from the Cauchy problem depend continuously on the values of
initial data--conditions which are consistent with a well-posed initial value problem. 
The particular procedure  adopted in this approach
borrows from the discussion of the evolution equations for the
Einstein-Euler system in
\citet{Fri98,FriRen00,Fri96,Reu98}, where a Lagrangian gauge  was used (flow lines
of the fluid  fix the preferred time).
However, there are  in general different ways of dealing with the issue of Cauchy problem and stability especially when it concerns self-gravitating systems  implying perturbation of the background. It is clear that in this special GR-MHD system the central
equation is  the {Bianchi identity}, intrinsically providing
 Weyl
tensor evolution equations.
This is borrowed from the
analysis of the conformal Einstein-Maxwell system of \citep{Fri91}--(see also  for similar \citet{EllEls98}  frame
formalism equations, while
there are several formulations of the evolution equations of the
EME system  for the  numerical
computations especially for
the conditions of ideal magnetohydrodynamics--\citep{BauSha03,ShiSek05,EtiLiuSha10,Fon03}).
In \citet{BauSha03} it  makes use of an ADM formulation for the
Einstein part of the system, which is  known to have problems
concerning hyperbolicity--\citep{Fri96}.  This hyperbolicity of the Einstein part of the system in an ADM formulation has
been addressed in \citet{ShiSek05,EtiLiuSha10} using a
BSSN formulation (BSSN equations are first order in time and second
order in the spatial coordinates\footnote{The Baumgarte-Shapiro-Shibata-Nakamura (BSSN) formalism,  consists in a    modification of the ADM formalism of General Relativity (in  Hamiltonian formalism   not making possible    long term and stable numerical simulations). The modification   introduced in the  ADM equations  by the BSSN formalism includes  consists essentially in the introduction of the  auxiliary variables   (to be considered with the  extrinsic curvature and  the three metric). Such change has been  used for  (long term)  analysis of  linear  and  non-linear gravitational waves  or black holes pair collision and generally different physics of spinning and double \textbf{BH}  systems or neutron stars and merging of neutron stars, collapse of spinning stars leading to a \textbf{BH} solutions.}), and  satisfied  under certain additional conditions in vacuum  the {strong
hyperbolicity}-- \citep{GunGar06,BauSha03,Alc08}.  Within  further assumptions the local existence and uniqueness of solutions for  these   systems can be ensured--\citep{Ren08}. Nevertheless this is a  mixed order system, also hard to implement in non-vacuum regions,   for example in the
EME systems\footnote{We note that the first order formulations  certainly  represent a clear simplification in many contexts and    are  generally preferred in numerical
computations as well as in the analysis of the motion of isolated bodies. Nevertheless,  we mention  that in \citep{ChoFri06} the local existence result for isolated dust (charged and uncharged)  bodies
 based on a mixed order formulation has been provided.
While, in \citet{Cho65,Cho08} the evolution equations for the EME system were
first considered from the perspective from the Cauchy problem,  where non-linear wave equations for the
metric tensor describing the gravitational field  adopting
 wave coordinates are obtained.  However as consequence of this choice  the evolution system is of
mixed order (in \citet{Cho08} where {Leray theory}  was implemented).  More general discussions of the problem of
the well-posedness of the evolution equations of the
EME system and of MHD can be found
in \citet{Put91,Fri74,Ren11,Put98,Zen03,ChoYor02}. Then,  the  initial
boundary value problem is a further relevant issue to be discussed, for example for the class of   maximally
dissipative boundary conditions used in \citet{FriNag99}, to show the
well-posedness of the initial boundary problem for the vacuum Einstein
field equations. }.

The procedure  should   ensure   the consistency of possible  numerical approaches to the problems. On the other hand,   the problem of the propagation of the constraints
can be   assumed  satisfied at
all time and then prove this result by fairly general arguments
as discussed in \citet{Reu98,Reu99}. Moreover,  the covariant and gauge--invariant
perturbation formalism   can be  more conveniently adapted to  spacetimes with some preferred spatial direction and possibly  applied
  to the case of gravitational wave propagation by
introducing a radial unit
vector and   decomposing all covariant quantities with
respect
to this \citep{Clarkson07,Marklund:2004qz,BauSha03}.
(In any case,  for the linear
perturbation analysis it is often useful, as done  in \citet{first,second},   to introduce a re-parameterized
set of evolution equations based of a suitable combination of  the
radially projected shear and  expansion. The resulting evolution system
was  then used in \citet{second} to  analyse the stability  problem for small linear perturbations of the background, thus the  re-parameterized set of evolution equations
can be collected in a proper symmetric hyperbolic form on which   the
perturbation to the first order of the variables can be discussed.).
The set of evolution equations has to be complemented by the
constraint and constitutive equations.  One can consider
 isotropic fluids (where entropy is a constant of both space and time) and one species particle fluid (simple fluid) and at later steps of the analysis one can
introduce a polytropic equation of state with a constant velocity of
sound.
To close the system of evolution equations it is
necessary to specify the form of the conduction current.  Accordingly,  in \citet{first,second}  the  Ohm's law was assumed so to obtain a linear relation between the conduction current
and the electric field, with a constant electrical conductivity
coefficient. Ultimately the pressure of the fluid $p$
and the charge density $\varrho_C$  can be assumed to be general functions of the matter
density $\rho$, and the charge current $j_c$ a function of the electric
field $E$.

Thus, summarizing ,
the {GR-MHD} system can  then be casted in quasi linear symmetric hyperbolic evolution system by prescribing suitable
initial data on an initial hypersurface.  We  focus on the
properties of the evolution system\footnote{We shall see in Sec.\il(\ref{Sec:Stationar}) and (\ref{Sec:influence})  how, in the case of gravitating toroids, using some symmetries of the background and the matter and fields adapted to the  geometry background and the  accretion orbiting disks,  the evolution equations are automatically satisfied leading to  the development of a constrain model for the constraining equations for the matter capable to provide a good approximation for the  analysis of a variety of orbiting accreting disks.}.and the analysis of its linear
stability,  while for  further aspects  of the propagation of the  constraints
we refer to   \citet{first}  and  in particular \citet{Reu98,Reu99}.
Consequently,   following \citet{first}  we can  write the  evolution equations for the
independent components of $n$ variables collected in  the $n$-dimensional
vector $\textbf{{v}}$, introduced  to obtain a suitable symmetric hyperbolic
evolution system of the form
\bea
\label{E:IIa}
\mathbf{A}^0\partial_0 \textbf{{v}}-\mathbf{A}^j\partial_j \textbf{{v}}=\mathbf{B} \mathbf{v}.
\eea
Quantities
in Eq.\il(\ref{E:IIa}) $\mathbf{A}^\mu=\mathbf{A}^0(x^{\mu},{\mathbf{v}})$ are matrix-valued
functions of the coordinates and the unknowns ${\mathbf{v}}$:  $\mathbf{A}^0$ and $\mathbf{A}^j$ and $\textbf{B}$ are smooth
matrix valued functions of the coordinates $(t, x)$ and the variables
$\textbf{{v}}$ with the index $j$ associated to some spatial
coordinates $x$.  The system is   symmetric hyperbolic if the
matrices $\mathbf{A}^0$ and $\mathbf{A}^j$ are symmetric and if
$\mathbf{A}^0$ is a positive-definite matrix. These evolution equations
must be  then complemented by constraint and  the constitutive
  equations.

\medskip

\textbf{Stability analysis for  systems with a privileged direction of symmetry}

In presence of special symmetries as the spherical symmetry  the problem can be addressed  by means of an adapted $(1+1+2)$ decomposition. In the first step of a  $(1+3)$  decomposition  the various tensorial quantities and equations are
projected along the direction of the  observer  comoving with the  fluid and onto its
orthogonal subspace. Quite often it is convenient   to assume,  to fix certain components of the connection especially in the self-gravitating systems,  the timelike vector of the defined orthonormal frame to follow the matter flow
lines (Lagrangian gauge) and also the vector fields
tetrad to be Fermi transported in the direction of $U$--\citep{Fri98,FriRen00}.
In \citet{first} a key feature of the  analysis was the introduction
of a further  tensor $\psi_{abc}=\nabla_a F_{bc}$, where $F_{ab}$ is the Faraday tensor and  $\nabla_a$ is the covariant
derivative. This implies to write the evolution and constraining equations for the 3-rank tensor $\psi$ ensuring however  the symmetric hyperbolicity of the
propagation equations for the components of the Weyl tensor (which is actually decomposed
into  its electric and magnetic
parts).

As we mentioned above this formalism and particularly the procedure adopted in \citet{first}, providing  the introduction of an auxiliary variable is particularly well suited to the study of the stability of more varied {GR-MHD} systems and including   configurations characterized   by some particular symmetries.
Considered as an additional exemplification,  we can take advantage of  the presence of a privileged direction of symmetry  in addressing the initial data problem and the stability  analysis as well:
the  initial value formulation of  {GR-MHD} discussed above
can be used  to discuss the stability of   configurations characterized by certain symmetries, for example in
spherically symmetric configurations.
This approach however breaks  the covariance of the
equations by introducing coordinates and choosing a preferred timelike
direction in the spacetime (thus prescribing   a gauge choice). In fluid models
 the convenient natural timelike direction  is given by the fluid flow-lines.
The motivation behind the gauge choice  is to obtain a
closed (as many equations as unknowns) subsystem of evolution equations  in
{symmetric hyperbolic} form
(this   procedure  is known as the  {hyperbolic reduction})--\citep{FriRen00}.
In this context, in \citet{second}
by means of a $(1+1+2)$-tetrad formalism
   {the linear stability of an }   Einstein-Maxwell perfect
fluid configuration  with a privileged direction of symmetry was examined.
The nonlinear stability for the case of an infinitely conducting plasma was also considered.
This system can  model  a self-gravitating ball  of Maxwell  matter fluid (stability problem for locally rotationally symmetric (LRS) solutions). More specifically,
the  analysis was restricted   to isotropic
fluid configurations and it was  assumed a   constant electrical
conductivity coefficient  $\sigma_J$ for the fluid (plasma).  Consequently to these assumptions, the  equations describing the evolution of the system were casted in  {quasi} linear symmetric hyperbolic form.

A further interesting aspect of the application  to this special system, is that the  threshold for the emergence of the instability appears in both contracting and expanding systems (according to  the kinematic expansion scalar $ \Theta $). The system instabilities enlighten  two ranges for the   density of matter, and shear scalar  $\Sigma$ along the privileged direction of the system.
In conjunction with this setup, including  the magnetic field but not dissipative effects, the stability analysis  proved  the instability emergence   with unstable modes  threshold, regulated  by the conductivity  parameter $\sigma_J$ and the speed of sound $v_s$.  The stability conditions
were strongly determined by the constitutive equations, the square of the velocity of sound and the electric
conductivity,
hence  a complete
classification and characterization  of various stable and unstable
 configurations followed. (We note that the special case of an infinitely conducting plasma
describes an adiabatic flow so that the entropy per particle is
conserved along the flow lines--for a more specific discussion on the stability of spherically symmetric plasma  see for example   \citet{Ho60,Las-Lun07,VCxL97,Gu-Sha1999}.).
The velocity of sound and the  conductivity  act in a different way for the systems according to different assumptions.
Contracting  systems  with a  fast contraction rate or  expansion according to this threshold are certainly unstable. In the other cases  instead the threshold  is provided by the density values $\rho$,  and the couple $(\Sigma, \Theta)$  in different ranges depending on    $(\rho, v_s, \sigma_J)$.
  Then, a further aspect to underline in this context is that, as proved in \citet{second}, the magnetic field does not have a specific role in determining the stability of the system, and  the  Maxwell field and the geometrical  effects   enclosed by the magnetic and electric  part of the Weyl tensor.
However, often in the handling of system equations,  further simplifications are due.   This is because   even  in simplest case the perturbed equations for the radially
projected acceleration $\la$ is  a rather complicated
expression of the other variables and  derivatives. Thus the necessity of a simplified form,
 grounded on  some assumptions on the configuration (for example assuming a
null radial acceleration for the reference solution, one can then fix the
expansion of the 3-sheets and 2-sheets, the radial part of the shear
of the 3-sheet, the twisting of the 2-sheet and the radial part of the
vorticity of the 3-sheet, etc), leading to  consider systems from different stability classes, where  stability conditions bind mainly the expansion or  contraction along the preferred direction with respect to different regimes  of the radial shears.

  First a  symmetric hyperbolic system was written. The simplest example of a configuration with LRS symmetry  is the spherical symmetric   configurations. The  assumption  on the local symmetry of the system  leads to a covariant
decomposition of  Einstein-Maxwell perfect fluid field equations
extending the usual $(1+3)$-formalism to a convenient adapted $((1+1+2))--$formalism.   In the $((1+1+2))$-formalism,  it is   assumed a further
(spatial) vector field $n^a$. This assumption reduced the   equations on the plane parallel
and orthogonal to $n^a$,  fixing  the spatial vector on the privileged symmetry direction (in \citet{second}  being the radial direction) at
each point of  LRS classes II space-times as described in \citet{Clarkson07} see also \citet{Las-Lun07,Clarkson07,Bets-Clark04,Bur-cqg-08,Burston:2007wt,Burston:2007ws} and  for a deeper and more general discussion about the MHD configurations  in spherical symmetry, see  \citet{Ho60,Las-Lun07,VCxL97,Gu-Sha1999}.
It is clear that for  spherically symmetric
spacetimes   (LSS),   $n^a$ can naturally  be taken to
 point in the radial direction of the spherical symmetry, which is  therefore the system privileged direction of the system  admitting a
one--dimensional isotropy group.
We precise that  the covariantly split in $(1+1+2)$ procedure leads to  scalars, vectors, and
transverse-traceless 2-tensors, with respect to $n^a$--
\citep{Clarkson07,Bets-Clark04}.
As we have mentioned earlier  this formalism can be equally used for the  analysis of the stability problem for the self-gravitating systems in
\citet{Las-Lun07} and  the same formalism has been used to analyse self
gravitating spherically symmetric charged perfect fluid
configurations in hydrostatic equilibrium.
One can consider the  perturbations of the  background or otherwise to fix the spacetime reducing  the self--gravitating to an orbiting configuration on a fixed background, as we do in Sec.\il(\ref{Sec:Stationar}) and Sec.\il(\ref{Sec:influence}) for toroidal configurations.
In \citet{second}  the more general case was addressed considering perturbations of the  gravitational part, in terms of scalars  of the Weyl tensor, considering  the symmetries   preserved by the perturbation and   assuming a constant velocity of sound and conductivity parameter.
Considering the full   gravito-electromagnetic  effects,  one can classify the  solutions in terms of the scalars of  the Weyl's conformal tensor.
The gravitational background  must, in any case, have the same symmetries of LRS systems.
In the approaches considered in \citet{second,first}  the electromagnetic fields are considered through real vector functions\footnote{We mention  here however that  in many of the LSR spacetimes
was naturally used  the complex variable  $\psi=E+i B$ and $\psi^*=E-i B$,  to
decouple the equations with the  appropriate symmetries and  obtain linear equations in the fields-- \cite{Bets-Clark04,
Bur-cqg-08,
Burston:2007wt,
Burston:2007ws}.}.

\medskip

\textbf{On gauge conditions
and techniques on stability analysis}

The   hyperbolic reduction  developed for the general system in \citet{LubbeKroon2011kz,first}  is
independent of geometric
gauge considerations.  In  a LRS  spacetime any  background quantities are scalars, implying   that  the  vector and tensor quantities are automatically gauge invariant, under linear perturbations
as a consequence of the Stewart-Walker lemma \citep{proc,Clarkson07}.
A number of simplified subsystems can be analyzed. (For example, in \citet{second}
background configurations with a vanishing radial acceleration
for the reference solution are considered, then exploring   models with
fixed values of particularly kinematic scalar, as the
$(1+1+2)$-projected expansion, shear, twisting and the vorticity of the
system).
As done in \citet{first}, to  provide a suitable
propagation equation used  in \citet{second},  an auxiliary  field  is used  to rephrase    the stability  problem for small nonlinear
perturbations of a background solution, for the fluid radial acceleration. This field corresponds
to the derivative of the matter density projected along the radial
direction.
Then, a first order perturbation to $\mathbf{v}$ of the form
$\mathbf{v}\mapsto \epsilon \mathring{\mathbf{v}}+
\breve{\mathbf{v}}$ is assumed,  where the parameter $\epsilon$ sets  here the order
of the perturbation and  $\breve{\mathbf{v}}$ describes  the (linear)
perturbation of  the background  solution.  Assuming the
background variables $\mathring{\mathbf{v}}$ to satisfy  the unperturbed system,
 an evolution system for the perturbations of the form
$
\mathring{\mathbf{A}}^t\partial_t \breve{\textbf{{v}}}-\mathring{\mathbf{A}}^j\partial_j \breve{\textbf{{v}}}=\mathring{\mathbf{B}} \breve{\mathbf{v}}$ was  obtained.
The core of the stability analysis,  adapted from discussion in \citet{Reu99,AlhMenVal10}, consisted of the study  of the background term $\mathring{\mathbf{B}} $ using some
relaxed stability eigenvalue conditions. (This procedure  could be used as  first step towards the  analysis of
{non-linear} stability, under suitable
circumstances).
Then indirect methods based  for example on  the inspection of the
characteristic polynomial can be used for the analysis of
the eigenvalues. (In other words  we  study  of the non principal  part of the matrix $\mathring{\mathbf{B}} $ of the  system using some
relaxed stability eigenvalue conditions, establishing conditions where the linear
   instability occurs\footnote{First simple method can be for example the
   Descartes criterion  to determine the {maximum} number of
   positive and negative real roots of the  characteristic
   polynomial and in particular  simple cases one can make use of the
   so-called Routh-Hurwitz criterion to determine the number of roots
   with positive and negative real part of the polynomial by
   constructing the Routh associated matrix.}.).
   The reason for adopting indirect methods to  explore the  system stability is that
   the elements of matrix
$\mathring{\mathbf{B}} $ are, in general, functions of the  space and
time coordinates and then adding a further complication to the analysis.
\subsection{Problem set-up: the equations}\label{Sec:prez}
Below we  introduce the basic equations describing a
relativistic charged
perfect fluid coupled gravity,  constituted by the EME
system.

For reference we report below a summary of main notations and convection introduced along this discussion.

\medskip

\textbf{Remarks on notation and convection}

The implementation of the $1 + 3$-formalism used in the present
article follows the notation and conventions of
\citet{first,second}. The 4-dimensional metrics $g_{ab}$ has
signature $(-,+,+,+)$. {When more convenient, we also introduce a multiplicative constant $\epsilon=\pm$ in leaving the  (pseudo-Riemannian) signature indeterminate in the first part of this review, precisely in   Sec.\il(\ref{Sec:ideal}) where we will deal with certain general  aspects of the {GR-MHD} equations set without fixing the metric, therefore preferring  an independent signature treatment. In the second part, Sec.\il(\ref{Sec:Stationar})  and Sec.\il(\ref{Sec:influence}) we will fix a precise geometric  background, writing  the  line element  for the geometry and   fixing  the signatures. }
The {Latin indices} $a, b, c...$ will denote
spacetime tensorial indices taking the values $(0,1,2,3)$ while
$i,j,k...$ will correspond to spatial frame indices ranging over $(1,
2, 3)$. The Levi-Civita covariant derivative of $g_{ab}$ will be
denoted by $\nabla_a$. Whenever convenient, we use the semicolon
notation. As usual, one has that $\nabla_a g_{bc}=g_{bc;a}=0$.

In what follows, the timelike vector field ({flow vector}) $U^a$
will describe the normalised future directed 4-velocity of the
fluid. Fixing the signature such that $U^a U_a=-1$ (indices are raised and lowered with
$g_{ab}$), the tensor $h^{ab}$ is the
projector onto the three dimensional subspace orthogonal to $U^a$,
thus, one has that
\(
h^a{}_{b}=\delta^a{}_{b}+ U^a U_b,\quad h^b{}_ah^a{}_c=h^b{}_c,\quad h^a{}_b h^b_a=3, \quad h^a{}_bU_a=0.
\)
Following the standard approach of $(1+3)$-formalisms, we split the
first covariant derivative of $U^b$ as
\be\label{E:deco3+1u}
\nabla_{a}U_{b}= \sigma_{ab}+ \omega_{ab}+ {\frac{1}{3}}\,\Theta
h_{ab}- \dot{U}_{b}U_{a},
\ee
where $\sigma_{ab}\equiv D_{\langle b}U_{a\rangle}$ with
$\sigma_{ab}U^b=0$, $\omega_{ab}\equiv D_{[a}U_{b]}$ with
$\omega_{ab}U^b=0$, $\Theta\equiv D^{a}U_{a}$ and
$\dot{U}_{a}=U^{b}\nabla_{b}U_{a}$ are, respectively, the {shear}
and the {vorticity} tensors, the {volume expansion scalar},
and the {4--acceleration} vector. We introduce the
vorticity vector $\omega^a\equiv\epsilon^{abc}\omega_{bc}/2$ where
$\epsilon_{abc}=U^d\epsilon_{dabc}$, $\epsilon_{abe}\epsilon^{abe}=6$
and $\epsilon_{dabc}$ stands for the totally antisymmetric tensor with
$\epsilon_{0123}=\sqrt{-\det{g_{ab}}}$. Then,
$\sigma_{ab}U^{a}=0=\omega_{ab}U^{a}=\dot{U}_{a}U^{a}$ by
construction. In the above expressions, the operator $D_a$ corresponds
to the 3-dimensional covariant derivative obtained from projecting the
spacetime covariant derivative in the distribution orthogonal to
$U_b$, for a generic 2-rank tensor
$T_{bc}$, one has that $D_aT_{bc}=h^s{}_a h^t{}_b h^p{}_c T_{st;p}$, and
$T_{st;p}=\nabla_pT_{st}$. For clarity, whenever necessary, projected
indices of a tensor will be highlighted by $\langle
\rangle$- brackets. For example, we write $T_{\langle a b \rangle} = h_a{}^c
h_b{}^d T_{cd}$. {Where $\dot{w}_{\langle a \rangle}\equiv h_a{}^b \dot{w}_b$ denotes
the directional covariant derivative along the flow vector
(\emph{Fermi derivative}) and $D_aw_b\equiv h_a{}^u
h_b{}^v\nabla_uw_v$ is the orthogonally projected covariant derivative
of a vector.}

\begin{description}
\item[Einstein-Euler-Maxwell equations: ]

We start by considering the following {Einstein equations}
\bea
G_{a b}= T_{ab}=T^{(f)}_{a b}+
T^{(em)}_{a b}
 \eea
 the  {energy-momentum tensor}
$T_{ab}$ singles out  the electromagnetic part $T^{(em)}_{a b}$ we discuss below  and  the  fluid momentum tensor $T^{(f)}_{a b}=(\rho +p) U_{a}
U_{b}-\epsilon\ p g_{a b}$
 for  the
{ideal fluid}, while $\rho$ and $p$ are, respectively, the {total
energy density} and {pressure} as measured by an observer moving
with the fluid.  The time-like vector field ({flow vector}) $U^a$
denotes the normalized future directed 4-velocity of the fluid. It
satisfies $U^a U_a=\epsilon$.  Associated to $U^a$
we introduce the {projection tensor}
$
h_{ab}\equiv g_{ab}-\epsilon U_a U_b,
$
projecting onto the three dimensional subspace orthogonal to
$U^a$. Indices are raised and lowered with $g_{ab}$. Thus, one has
that $h^a{}_{b}=\delta^a{}_{b}-\epsilon U^a U_b$,
$h^b{}_ah^a{}_c=h^b{}_c$, and $h^a{}_bU_a=0$.
\\
\item[{The electromagnetic energy-momentum tensor}: ]

The tensor $T^{(em)}_{a b}$ denotes the energy momentum tensor of
an electromagnetic field:
 \be\label{E:ff}
  T^{(em)}_{a
b}=-\epsilon\left (F_{a c}F^{\phantom\ c}_{b}-\frac{1}{4} F_{c d} F^{c
d} g_{ab}\right),
\ee where $F_{ab}=\epsilon( 2 E_{[a}U_{b]}
-\epsilon_{abcd}B^{c}U^{d})  $ is the {electromagnetic
field (Faraday) tensor}. The latter can be split in its {electric
part}, $E_{a}\equiv F_{ab}U^{b}$, and its {magnetic part},
$B^{a}\equiv\frac{1}{2}\epsilon^{abcd}U_{b}F_{cd}$, with respect to
the flow.
One can readily verify the properties
$E_{a}U^a=B_aU^a=0$. Using the decomposition into electric and
magnetic parts, the electromagnetic energy-momentum tensor,
Eq.\il(\ref{E:ff}), can be written as
\bea\nonumber
&&
T^{(em)}_{a b}\equiv-\frac{\epsilon}{2} U_aU_b (E^2+B^2)+\frac{h_{ab}}{6}(E^2+B^2)+P_{ab}-2\epsilon \mathcal{G}_{(a}U_{b)},\quad (E^2\equiv E_aE^a),\quad (B^2 \equiv B^aB_a)
\\\label{E:ff2}
&&P_{ab}=P_{(ab)}\equiv\frac{h_{ab}}{3}(E^2+B^2)-(E_aE_b+B_aB_b), \quad \mathcal{G}_{a}\equiv\epsilon_{auvd}E^uB^vU^d,\eea
where $P_{ab}$ is  symmetric, trace-free tensor and $\mathcal{G}_a$
denotes the {Poynting vector}.
\\
\item[The Maxwell equations: ]

The Maxwell equations are given by
\be\label{E:MW}
\nabla_{[a}F_{b c]}=0,\quad\nabla^a F_{ab}=\epsilon J_b.
\ee
%
where $\dot{E}\equiv U^a\nabla_a E_b$ stands for the covariant
derivative of $E_a$ along the flow ($\nabla_{a}(E^bU_b)=E_b\nabla_aU^b+U_b\nabla_aE^b=0$). 
Projecting along the directions parallel and
orthogonal to the flow vector $U^b$ one obtains the  two equations
where the electromagnetic current vector $J^{a}$ will be split with respect
to the flow vector as
$
J^a=\rc U^a +j^a,
$
where $\rho_{\ti{C}}$ denotes the {charge density} and $j^a$ is
the {orthogonally projected conduction current}. The  propagation equations for the
electric and magnetic parts of the Faraday tensor  and  the constraint equations are
\bea\label{E:UE2a}
&&
\dot{E}_{\langle f \rangle}=-2E^a h_{f[a}\nabla_{b]} U^b-\epsilon_{abcd}h^b{}_f\nabla^a(B^c U^d)-j_f,
 \\\nonumber
 &&\epsilon D^aE_a=\epsilon_{abcd}U^a B^b\nabla^c U^d+\epsilon \rho_{\ti{C}},
\\&&
\label{E:Ub2a}
\dot{B}_{\langle f \rangle}=-2B^a h_{f[a}\nabla_{b]} U^b+\epsilon_{abcd}h^b{}_f\nabla^a(E^cU^d),\\
 &&\nonumber\epsilon D^aB_a=-\epsilon_{abcd}U^bE^c\nabla^a U^d,
\eea
It can  be necessary to specify the form of the
conduction current, $j^a$ and, consistently  with {Ohm's law}, one can  assume
a linear relation between the conduction current $j^a$ and
the electric field as
$
j^a=\sigma^{ab}E_b,
$
introducing the   {conductivity} of the fluid
(plasma) $\sigma^{ab}$. In many applications we can restrict  the analysis  to isotropic fluids  where
$\sigma^{ab}=\sigma g^{ab}$,   implying
$
J^a=\rc U^a +\sigma E^a,
$
with $\sigma$ the {electrical conductivity coefficient}.
A special case is represented by the ideal
\textbf{MHD} which is  characterized by the condition, $\sigma\rightarrow\infty$
(i.e. an ideal conductive plasma), implying thus  the constraint $E_a=0$.
\\
\item[The fluid equations: ]

Form  the conservation of the energy--momentum tensor $\nabla^a T_{ab}=0$  we find
\bea\label{:B}&&
 \nabla^{a}T^{(em)}_{ab}=-\epsilon\left(\nabla^a F_{ac}\right)F_{b}^{\phantom\ c},
 \\\nonumber
  &&\nabla^{a}T^{(f)}_{ab}=U_bU_a\nabla^a (p+\rho)+(p+\rho)\left[U_b(\nabla^aU_a)+U_a \nabla^a U_b\right]-\epsilon\nabla_b p.
\eea
Considering the projections
along the directions parallel and orthogonal to the flow lines  $U^a$ of the
fluid we obtain the
conservation equations and the Euler (acceleration) equation
\bea\label{Eq:conservazione}&&
U_a\nabla^a\rho+(p+\rho)\nabla^aU_a- U^bF_{b}^{\phantom\ c}(\nabla^aF_{ac})=0,
\\
&&\nonumber(p+\rho)U^a\nabla_aU^c-\epsilon h^{bc}\nabla_b p-\epsilon(\nabla^aF_{ad})F^{\phantom\ d}_b h^{bc}=0.
\eea
 (For an ideally
conducting fluid, there is  $E_a=F_{ab}U^{b}=0$, and the last term of
Eq.\il(\ref{Eq:conservazione}) is identically zero, thus
electromagnetic field does not have a direct effect on the
conservation equation along the flow lines). The {Euler equation}  is obtained by contracting (\ref{:B}) with the projector $h^{bc}$, where
in the ideal MHD case   it  reduces to
\be
(p+\rho)U^a\nabla_aU^c-\epsilon h^{bc}\nabla_b p-\epsilon(\nabla^aF_{ad})F^{cd}=0.
\ee
\end{description}
In \citet{first}  evolution equations have been constructed containing a part related to the gravitational field perturbation
  provided for example by the  (trace-free) the electric and magnetic parts of the
Weyl tensor--\citep{FriRen00}.

\medskip

Evolution equations can  then be  written for  the {independent} components of the vector
variable
${v}=\left(e_i{}^a,\g{0}_{0\ i},\g{a}_{i\
b},\hat{E}_{ab},\hat{B}_{ab},E_{a},B_{a},\mathcal{E}_{a},
\mathcal{B}_{a}, n, s , s_a\right)$.
The propagation equation for the tetrad coefficients,
$e_i{}^a$, are not included in this discussion as well as
the evolution equations for the electric part,
$\hat{E}_{ab}$, and the magnetic part, $\hat{B}_{ab}$, of the Weyl
tensor, the evolution equations for the electric part,
$\mathcal{E}_a$, and magnetic part, $\mathcal{B}_b$, of the field
$\psi_{abc}$, corresponding to the covariant derivative of the Faraday
tensor.
We remind to the extended discussion in \citet{first}. ($\g{a}_{b\ c}$ are
connection coefficients--Ricci rotation coefficients, of  the tetrad $ e_{a}$,  defined by the relations
$g_{ab}=\eta_{ab}\omega^{a}{}_\alpha\omega^{b}{}_\beta$, $
g^{ab}=\eta^{ab}e_{a}{}^\alpha e_{b}{}^\beta$, where
$\nabla_{a}e_{b}=\g{c}_{a\ b}e_c$ and
$\nabla_{a}\omega^{b}=-\g{b}_{a\ c}\omega^c$.
Since
$e_{a}\left(\eta_{bc}\right)=0$, there is the symmetry
$\Gamma_{a(bc)}=0$.)
The propagation equation for the tetrad coefficients,
$e_i{}^a$,  give a symmetric hyperbolic subsystem of equations.
The equations for the connection coefficients, $\g{0}_{0\ i}$ and $\g{a}_{i\
b}$ are given within  gauge condition, $\Gamma_{0\
i}^{\phantom\ j }=0$. (Components of connections terms $\Gamma_i{}^j{}_k$  are ruled out  by means of gauge conditions and
symmetries.).
To these must be added the terms  from the matter and field component, while the former equations for the tetrads and the background  may be simply disregarded  for the  gravitating systems on the fixed background geometry.
More precisely we consider:
\begin{description}
\item[
--] The evolution equations for the particle number density $n$  (we can use a Lagrangian gauge where all the derivatives
of the flow vector $U^a$ are replaced by the  connection coefficients, and
thus, it contains no derivatives in the spatial directions).
\item[
--] The evolution equation for the entropy per particle, $s$, which in general depends   on the auxiliary field
$\psi_{abc}$, i.e. the derivative of the Faraday tensor,  or alternately,  by means of the
inhomogeneous Maxwell equation, on the   current vector $J_a$.
\item[
--] Finally the equation for the vector  $s_a\equiv\nabla_a s$.
\end{description}
\subsection{Thermodynamic considerations}\label{Sec:terms-cons}
\begin{itemize}
\item[-] \textbf{General considerations on the first law of thermodynamics}

In this paper  we consider a {one species particle fluid} (simple
fluid). With  $(n,\,s,\, T)$ as the {particle number density},
the {entropy \emph{per} particle} and the {absolute temperature} respectively
as measured by comoving observers, we  introduce the {volume $v\equiv\frac{1}{n}$
\emph{per} particle} and the {energy \emph{per} particle}, $e\equiv \frac{\rho}{n}$.
 First law of thermodynamics
$\mbox{d}e=-p\mbox{d}v+T\mbox{d}s$,  can be therefore  written  in terms of these variables as follows
 \be\label{E:1law}
\mbox{d}\rho=\frac{p+\rho}{n} \mbox{d}n +n T \mbox{d}s.
%
\quad
p(n,s)=n\left(\frac{\partial \rho}{\partial n}\right)_{s}-\rho(n,s),
\quad\mbox{where}\quad T(n,s)=\frac{1}{\rho}\left(\frac{\partial \rho}{\partial s}\right)_n.
\ee
where in the second and third equalities  we assumed an equation of state of the form $\rho=f(n,s)\geq0$. If we also assume that $\partial p/\partial \rho>0$ and  define the {speed
of sound}, $\nu_s=\nu_s(n,s)$, thus there is  $
\nu_s^2\equiv\left(\frac{\partial p}{\partial \rho}\right)_s=\frac{n}{\rho+p}\frac{\partial p}{\partial n}>0.
$.

The absence of  particle creation  or annihilation is translated into an equation of conservation of particle number:
$
U^a\nabla_a n+n \nabla_a U^a=0
$,
which with  Eqs.\il(\ref{Eq:conservazione}) and \il(\ref{E:1law}) we obtain
\be\label{E:fors}
U^a\nabla_a s=\frac{1}{nT}U^b F_{b}^{\phantom\ c}\nabla^a F_{ac}.
\ee
In the case of an infinitely conducting plasma, where the last term of
Eq.\il(\ref{E:fors}) vanishes, Eq.\il(\ref{E:fors}) describes an
adiabatic flow ---that is, $U^a\nabla_as=0$, so that the entropy per
particle is conserved along the flow lines. A particular case of
interest is when $s$ is a constant of both space and time.  In this
case the equation of state can be given in the form $p=p(\rho)$.
\\
\item[-]\textbf{{Homentropic flows}}
A fluid is said to be \emph{homeontropic} if  $s$ is constant in space
and time. In general, the equation of state is given by $\rho=
f(n,s)$. Now, if one has an homeontropic flow, the latter can be
written as $\rho=f(n)$ ---there is no dependence on $s$ as it is
constant. Now, if $f$ is a differentiable function of $n$ and
$f'(n)\neq 0$, then one can write $n =f^{-1}(\rho)$. Thus, $p$ is of the form $p=h(\rho)$ ---that is, one obtains a
barotropic equation of state. In fact there is
that
\begin{eqnarray}
&& p = n \frac{\partial \rho}{\partial n} - \rho= n f'(n) - \rho= n f'(f^{-1}(\rho))-\rho.
\end{eqnarray}
\\
\item[-]
\textbf{Conditions for a barotropic equation of state}
Thus the assumption of an
homeontropic flow implies a barotropic equation of state. We now
investigate more general situations for which one can have this type
of equation of state.
Assume that $p=h(p)$ . Then, from
\bea
p(n,s) = n \left( \frac{\partial \rho }{\partial n}\right) - \rho(n,s)
\quad \mbox{thus}\quad
h( f(n,s)) = n \left( \frac{\partial f }{\partial n}\right) - f(n,s).
\eea
The latter can be read as a (possibly non-linear) differential
equation  for $f(n,s)$. It can be integrated to obtain
$
\exp\left( \int \frac{\mbox{df}}{h(f)+f}  \right) = g(s) n,
$
where $g(s)$ is an arbitrary function of $s$. This expression can be
used to eliminate $n$ from the discussion.  In the case of a dust
equation of state ($p=0$), the last relation reduces to
$
\rho/ n = g(s).$
\end{itemize}

\subsection{Infinitely conductive plasma}
\label{Subsection:MHD}
A particularly important subcase of our system  is that of
an ideal conductive plasma where  ideal
magnetohydrodynamics is implemented and condition  $E_a=0$  holds (an hyperbolic problem for this case can be naturally treated as a subcase
of the non ideal case. Note however that, even within   $E_a=0$,  there can be still a non-vanishing
electric part, $\mathcal{E}_i=-\g{j}_{a\ d}F_{ij}U^d$, of $\psi_{abc}$.

Before starting the second part of this article, it is convenient to  review  some basic notions of  ideal GR-MHD summarizing the analysis of Sec.\il(\ref{Sec:prez}) and Sec.\il(\ref{Sec:terms-cons}).  The  fluids energy-momentum tensor  can be written as  the composition  of the two components
\bea
\label{E:Tm}&& T^{(f)}_{a b}=(\rho +p) U_{a}
U_{b}-\epsilon\ p g_{a b}
\\\nonumber&& T^{(em)}_{a
b}=-\epsilon\left (F_{a c}F^{\phantom\ c}_{b}-\frac{1}{4} F_{c d} F^{c
d} g_{ab}\right)=
\\\nonumber
&&\frac{g_{ab}}{2}(E^2+B^2)-(E_aE_b+B_aB_b)-2\epsilon \mathcal{{G}}_{(a}U_{b)}-\epsilon U_aU_b(E^2+B^2)
\eea
The Maxwell (and constitutive) equations are
\bea
\nabla_{[a}F_{b c]}=0,\quad\nabla^a F_{ab}=\epsilon J_b\quad
J^a=\varrho_c U^a +j^a,
\eea
(all the quantities are  measured by  observers moving
with the fluid), here we remind $U^a U_a=\epsilon$,  ($\epsilon$ is clearly a signature sign) and therefore  the projection tensor $
h_{ab}\equiv g_{ab}-\epsilon U_a U_b,
$.
Considering the  {charge density} and  {conduction current} with the
  {Ohm's law},  there is
$
j^a=\sigma^{ab}E_b,\quad
J^a=\varrho_c U^a +\sigma E^a
$,
we consider isotropic fluids for which
$\sigma^{ab}=\sigma g^{ab}$,
$\sigma$ is the {electrical conductivity coefficient}. For ideal conductive plasma there is $\sigma\rightarrow\infty$ ($E_a=F_{ab}U^{b}=0$): the
electromagnetic field does not have a direct effect on the
conservation equation along the flow lines. This situation is described by the continuity and Euler equations below
\bea\nonumber\label{Eq:conservazione}
&&U_a\nabla^a\rho+(p+\rho)\nabla^aU_a- U^bF_{b}^{\phantom\ c}(\nabla^aF_{ac})=0,
\\\label{Eq:preg-t}
&&\mbox{In the ideal MHD}
\quad
(p+\rho)U^a\nabla_aU^c-\epsilon h^{bc}\nabla_b p-\epsilon(\nabla^aF_{ad})F^{cd}=0,\\
&&\nonumber\mbox{and}\quad
U^a\nabla_a s=\frac{1}{nT}U^b F_{b}^{\phantom\ c}\nabla^a F_{ac}.
\eea
Thus, in the infinitely conductive plasma case  the system of equations reduces to
\begin{eqnarray}
&& U_a\nabla^a\rho+(p+\rho)\nabla^aU_a=0, \label{E:1}
\\
&& (p+\rho)U^a\nabla_aU^c-\epsilon h^{bc}\nabla_b p-\epsilon(\nabla^aF_{ad})F^{\phantom\ d}_b h^{bc}=0,
\\\label{E:somo}
&& U^a\nabla_a s=0.
\end{eqnarray}
(Consistent with Ohm's law the source term of the
Maxwell equation is in this case simply given by $J_a=\rho_{\ti{C}}
U_a$).
Note that as a consequence of Eq.\il(\ref{E:somo}) the
entropy per particle is constant along the flow lines of the perfect
fluid, which is translated into  the propagation
equation for $s$ $\mathcal{L}_U
s_a=0$ ($\mathcal{L}_U$ is for the Lee derivative).

\section{The GR-HD systems: thick   tori  orbiting  Kerr SMBHs
}\label{Sec:Stationar}
Configurations of extended matter rotating around gravitational sources  are  an environment  with   rich inner  structure and  unstable phases. Their phenomenology covers very different aspects of High-Energy Astrophysics, from the proto-planetary accretion disks  to the  violent dynamical effects  such as Gamma Ray Bursts. Such processes, combined with  the presence of a very compact object as the central attractor,  can  have  extremely large radiative energy outputs and   result often in  ejection of matter associated with jet-like structures emerging from extremely small central regions, for example associated to the inner edge of accretion disks and more generally to accretion processes. All these different processes are attributed to different (mostly unclear) mechanisms empowered by  the strong gravity of a central \textbf{BH} attractor.
There are  therefore  several  models  for  orbiting  accretion disks, classified   according to several  characteristics as their   optical depth, accretion rates or  geometric structure.
 Accreting models  can be distinguished by  the geometry (the vertical thickness), accordingly  we can distinguish geometrically thin or thick disks;  the matter accretion rate (we can define sub- or super-Eddington luminosity), and the optical depth (i.e., transparent--optically thin, or opaque--optically thick disks), \citep{abrafra}.
Geometrically thin disks are modelled as the standard Shakura-Sunayev (Keplerian) disks \citep{[S73],[SS73],Page-Thorne74}. Geometrically thick disks are modelled  often as  Polish Doughnuts \citep{Koz-Jar-Abr:1978:ASTRA:,AJS78,Jaroszynski(1980),Stu-Sla-Hle:2000:ASTRA:,Rez-Zan-Fon:2003:ASTRA:,Sla-Stu:2005:CLAQG:,[68],mnras1,epl,pugtot}, or the ion tori \citet{Rees1982}.  The thick  and slim disks have high optical depth being  opaque, while the ion tori and the ADAF disks have low optical depth being thus transparent. For the ADAF (Advection-Dominated Accretion Flow) disks see for example  \citet{Ab-Ac-Schl14,Narayan:1998ft}, and  for the slim disks  the \citep{abrafra}.
The ADAF disks and the ion tori have relatively low accretion rates (sub-Eddington), while the  thick disks we consider here    have very high  accretion rates  (super-Eddington).

An accretion disk is essentially regulated by the balance of different factors as the  gravitational, centrifugal and magnetic components. At the same time, the dynamics of any accretion disk is determined by   centrifugal, dissipative, and magnetic effects.
 In the geometrically thin disks, dissipative viscosity processes are relevant for accretion, being usually attributed to the magnetorotational instability of the local magnetic fields \citep{Hawley1984,Hawley1987,Hawley1990,Hawley1991,DeVilliers}. In the toroidal disks, pressure gradients are crucial \citep{AJS78}.
 Thick disks  characterize  regions very close to the attractors,  requiring a  full general relativistic treatment.
 Geometrically thick accretion disks are strongly characterized by the  gravitational forces of the central super-massive attractors, consequently are characterized as among the most energetic  astrophysical objects  in environments as  the  Active Galactic Nuclei  (\textbf{AGN}) as seen as  engines empowering many  high energetics phenomena.
  In many accretion disks, gravitational force constitutes the basic ingredient of the accretion mechanism   independently of any dissipative effects that are   strategically   important   for the accretion processes in the thin models \citep{Balbus2011,[SS73]}.
 This model and its derivations are widely studied in the literature with both numerical and analytical methods, we refer  for  an extensive bibliography to \citet{abrafra}.
In  \citet{Komissarov:2006nz} and \citet{Montero:2007tc},  the fully relativistic theory of
stationary axisymmetric   torus in Kerr metric has been generalized   by including
a strong toroidal magnetic field,  leading to the  analytic solutions for barotropic tori \citep{epl,Fi-Ringed}. We discuss this special model  in Sec.\il(\ref{Sec:influence}).

In this work we consider  toroidal  (thick disk) configurations  centered on a  Kerr \textbf{BH}, and prescribed by  barotropic models, for which the time-scale of the dynamical processes $\tau_{dyn}$ (regulated by the gravitational and inertial forces,  balancing  the pressure and gravitational and centrifugal force) is much lower than the time-scale of the thermal ones $\tau_{therm}$   (heating, cooling processes and  radiation, in general  dissipative heating), that is lower than the time-scale of the viscous processes $\tau_{\nu}$, or $\tau_{dyn}\ll\tau_{therm}\ll\tau_{\nu}$. Consequently the effects of strong gravitational fields are generally dominant with respect to the  dissipative ones and predominant to determine  the systems unstable phases \citep{abrafra,pugtot}. On the contrary the plasma configuration instability, especially in the  geometrically thin  structures  orbiting around an attractor  (es Shakura-Sunyaev accretion disks)  is   described often by  the  magneto-rotational instability. The dissipative (visco-resistive) effects are essential in these models as they  allow the transport of angular momentum in the configurations in accretion on the central object.
In fact, in the geometrically thin configurations  it is assumed that the time scales of the  dynamical process  are less then  the  thermal one  that is less then the viscous ones which in these models comprise the timescales of  the  dissipative stresses leaded by  the consequent angular momentum transport inside the disk--\citep{Jaroszynski(1980),Pac-Wii,cc,Koz-Jar-Abr:1978:ASTRA:,abrafra}
see also \citet{F-D-02,abrafra,pugtot,Pac-Wii,Hawley1990,Fragile:2007dk,DeVilliers,Hawley1987,Hawley1991,Hawley1984,Fon03,Lei:2008ui}.
The magnetic field,  the dissipative effects and the radial gradient of the plasma relativistic angular velocity are therefore essential for the  MRI instability. However, some aspects of the theoretical framework of the  MRI and accretion process  are still to be clarified. An  intriguing  issue  for example is  the so-called visco-resistive puzzle: eventually high values of and resistivity  and  viscosity  have to be assumed.

Here instead we  consider the  situation where the  gravity  of the central attractor  is a predominant component of the disks force balance, requiring consequently    a full  general relativistic analysis for the disks, gravity   plays therefore  a decisive role in determining both the equilibrium states of the  configurations and  the dynamical  phases  associated  to the  instability. Consequently  the  torus  model  provides an
analytical description of the orbital  structure, considering  a perfect  fluid circularly orbiting  around  Kerr  background geometry,  and described by the effective potential approach for the exact gravitational and centrifugal
effects.
The model  leads to a  detailed, analytical description of the accretion
disk, its toroidal surface, the thickness, the  distance from the source.
In this case the  torus shape is defined by the constant  Paczynski-Wiita potential $W$.   In this approach
closed equipotential surfaces define stationary equilibrium configurations where  the fluid
can fill any closed surface. The torus thickness  increases with the
 ``energy'' parameter $({{K}})$, defined as the
value of the  potential for that momentum $\ell$, and decreases with the angular momentum: the torus becomes thicker  for high energies and
the { maximum diameter} $\lambda$ increases with ${K}$.
The open equipotential surfaces define
dynamical situations, for example the formation of matter jets \citep{Boy:1965:PCPS:,proto-jet,open}.
The critical, self-crossing and closed equipotential surfaces (cusp) locates the accretion onto the  central \textbf{BH} due to Paczy\'nski
mechanism. In this hydromechanical instability process,
a little overcoming of the critical equipotential surface corresponds to  a violation of the hydrostatic equilibrium when
  the disk surface exceeds
the critical equipotential surface.
Consequently the relativistic
Roche lobe overflow at the cusp of  the equipotential surfaces is also  the
 stabilizing mechanism against the thermal and viscous instabilities {locally},
and against the so called Papaloizou--Pringle instability {globally} \citep{Blaes1987}.
(For a discussion on the relation  between  Papaloizou-Pringle (\textbf{PP})  global incompressible modes  in the tori, the  Papaloizou-Pringle Instability
(\textbf{PPI}),
a global, hydrodynamic, non-axis-symmetric instability   and the  Magneto-Rotational Instability (MRI)  modes see  \cite{Fi-Ringed,Bugli}, we consider this issue more closely in Sec.\il(\ref{Sec:influence})).
This  fully general
relativistic model of   pressure supported  torus,
 traces back to the Boyer theory of     the equilibrium and rigidity in general relativity, i.e. the  analytic theory of equilibrium configurations of   rotating perfect fluids \citep{Boy:1965:PCPS:}.
Within the so  called ``Boyer's condition'', we   can  determine    the boundary of  stationary, barotropic, perfect fluid body as  the
surfaces of constant pressure (eventually also   equipotential surfaces). This   occurs in many models, essentially due to  the condition $\Omega=\Omega(\ell)$ on the fluid relativistic frequency $\Omega$   that has   to be a     function of  $\ell$ (fluid specific angular momentum), a result known as
von Zeipel condition \citep{Zanotti:2014haa,Koz-Jar-Abr:1978:ASTRA:,M.A.Abramowicz,Chakrabarti,Chakrabarti0}.
The {opaque} and
Super-Eddington disk we consider here is  parameterized by
an
ad hoc distribution of constant angular momentum.
 The choice $\ell=$constant for each torus is a well known assumption, widely used in several contexts where geometrically thick tori are considered--\citep{abrafra}.
The advantage of this model turns  to be both conceptual and technical.  From a technical view-point, essential   features of the disk  morphology  like the thickness, the elongation   on its symmetric (equatorial) plane, the distance from the attractor are predominantly regulated  by the geometric properties of spacetime via  the pressure gradients in  the relativistic Euler equation, reducible to an ordinary differential equation (ODE), often integrable with the introduction of an effective potential. In this context also the torus inner (outer) edge are well defined and constrained,  in the different torus  topological phases related to the stable and unstable  phases of the disk (emergence of the Roche lobe and  the cusp formation).
Lastly, these configurations have been often adopted as the initial conditions in the set up for simulations of the MHD  accretion structures\citep{Igumenshchev,Shafee} and \citep{Fragile:2007dk,DeVilliers}, resulting therefore a model of great applicability.

\medskip

In the following sections we develop the discussion as follows: in Sec.\il(\ref{Sec:kerr-flji}) we write  down the Kerr metric for the central attractor  establishing  the conventions  adopted throughout the second part of the work. In  Sec.\il(\ref{Sec:la-stafdisc-gtr}) we introduce the   {GR-HD}  tori orbiting around a Kerr \textbf{SMBH} introducing the main quantities, the key concepts related to disk model and the main notation.
In Sec.\il(\ref{Sec:constr}) we discuss constraints, morphology and angular momentum  distribution of the disk.
We conclude this section on {GR-HD} thick tori  reviewing aspects related to the tori instability, the  model construction and phenomenology  in the context  of multi-accretion processes, multi--orbiting structures  and ringed accretion disks-- Sec.\il(\ref{Sec:intro-RAD}).


%

 \subsection{The Kerr geometry}\label{Sec:kerr-flji}
 We start introducing  the Kerr geometry in Boyer-Lindquist (BL) coordinates
\bea\label{Eq:metric-1covector}
&&
\dd s^2=-\alpha^2 \dd t^2+\frac{A \sigma}{\rho^2} (\dd \phi- \omega_z \dd t)^2+\frac{\rho^2}{\Delta}\dd r^2+\Sigma \dd\theta^2,\quad\mbox{where}\quad\sigma\equiv\sin^2\theta,
\\
&&
 A\equiv (r^2+a^2)^2-a^2 \Delta \sigma\quad \Delta\equiv r^2-2Mr+a^2,\quad\mbox{and}\quad\Sigma\equiv r^2+a^2\cos^2\theta \ .
\eea
we introduced also
$\alpha=\sqrt{(\Delta \Sigma/A)}$  and $\omega_z=2 a M r/A$ as the lapse function\footnote{We adopt the
geometrical  units $c=1=G$ and  the $(-,+,+,+)$ signature, Latin indices run in $\{0,1,2,3\}$.  The   four-velocity  satisfy $U^a U_a=-1$. The radius $r$ has unit of
mass $[M]$, and the angular momentum  units of $[M]^2$, the velocities  $[U^t]=[U^r]=1$
and $[U^{\phi}]=[U^{\theta}]=[M]^{-1}$ with $[U^{\phi}/U^{t}]=[M]^{-1}$ and
$[U_{\phi}/U_{t}]=[M]$. For the seek of convenience, we always consider the
dimensionless  energy and effective potential $[V_{eff}]=1$ and an angular momentum per
unit of mass $[L]/[M]=[M]$.} and the frequency of the zero angular momentum fiducial observer \textbf{zamos} whose four velocity is $U^a=(1/\alpha,0,0,\omega_z/\alpha)$ orthogonal  to the surface   of constant $t$.
 The rotational parameter associated to  the central singularity is the spin (the  {specific} angular momentum) $a\equiv J/M $,  and  $M\geq0$  is      interpreted as  the mass  of the gravitational source, while   $J$ is the
{total} angular momentum. For  $a=0$, the metric  (\ref{Eq:metric-1covector})  describes   the limiting static and spherically symmetric  Schwarzschild  geometry. The limiting condition  $a=M$ is the extreme Kerr \textbf{BH} where there is $r_+=r_-=M$.
Alternately we can write the line element  Eq.\il(\ref{Eq:metric-1covector}) as follows
{\small{
\bea
\label{alai}
&& ds^2=-\frac{\Delta-a^2 \sin ^2\theta}{\Sigma}dt^2+\frac{\Sigma}{\Delta}dr^2+\Sigma
d\theta^2+
\\
&&\nonumber\frac{\sin^2\theta\left(\left(a^2+r^2\right)^2-a^2 \Delta \sin^2\theta\right)}{\Sigma}d\phi^2
-2\frac{a M \sin^2\theta \left(a^2-\Delta+r^2\right)}{\Sigma}d\phi dt.
\eea
}}
The inner and outer  horizons and the inner  and outer static limits (ergosurfaces) for the {Kerr} geometry  are,
$
r_{\mp}=M\mp\sqrt{M^2-a^2}$ and $ r_{\epsilon}^{\mp}=M\mp\sqrt{M^2-a^2 \cos ^2\theta}\ ,
$ respectively.
These  event horizons are    Killing horizons   with respect to  the Killing field
$\mathcal{L}_H=\partial_t +\omega_H^{\pm} \partial_{\phi}$, where  $\omega_H^{\pm}$ is the angular velocity {(frequency)} of the horizons  representing   the \textbf{BH} rigid rotation.
The vectors  $\xi_{t}=\partial_{t} $  and
$\xi_{\phi}=\partial_{\phi} $  are the stationary
  and axisymmetric  {Killing} fields,  respectively. %
$
{E} \equiv -g_{ab}\xi_{t}^{a} p^{b},$ and $ L \equiv
g_{ab}\xi_{\phi}^{a}p^{b}\ ,
$
are  constants of motion ($p^a$ stays for  the fluid four-momentum), associated with the circular geodesic motion relative to the orbits  of  the stationary
  and axisymmetric  {Killing}  vectors respectively.

\subsection{The HD-"stationary" configurations for thick tori}\label{Sec:la-stafdisc-gtr}

 The {GR-HD}  system we consider here,  is described by a perfect (simple) fluid  energy momentum tensor
 $T^{(f)}_{ab}$
of Eq.\il(\ref{:B}),
\bea
\label{E:II}&&
	T^{(f)}_{ab}\equiv T_{ab}=(p+\rho)U_a U_b - p g_{ab},
\eea
(in the HD model  the electromagnetic component is  null).
The  fluid dynamics  is described by the {continuity  equation} and the {Euler equation} respectively:
\bea\nonumber
&&
U^a\nabla_a\rho+(p+\rho)\nabla^a U_a=0\,
\quad
(p+\rho)U^a \nabla_a U^c+ \ h^{bc}\nabla_b p=0,
\eea
--see also Eq.\il(\ref{E:1}) and Eq.\il(\ref{E:somo})--
where the projection tensor $h_{ab}=g_{ab}+ U_a U_b$ and $\nabla_a g_{bc}=0$, (here $\rho$ and $p$ are  the {total fluid density} and
pressure, respectively, as measured by an observer moving with the fluid whose four-velocity $U^{a}$  is
a timelike flow vector field) \citep{pugtot,first}.
 Because of the symmetries of the system (stationarity and axial-symmetry), the orbiting configurations are  regulated  by the Euler equation only with a barotropic equation of state (\textbf{EoS}) $p=p(\rho)$.
More precisely, we always assume $\partial_t \mathbf{Q}=0$ and
$\partial_{\phi} \mathbf{Q}=0$, $\mathbf{Q}$  being a generic spacetime tensor investigating  the  fluid toroidal  configurations centered on  the equatorial   plane $\theta=\pi/2$ (of the central attractor and of the orbiting disk), and  defined by the constraint
$U^r=0$: no
motion is assumed  in the $\theta$ angular direction ($U^{\theta}=0$).
The  continuity equation
is  identically satisfied as consequence of these conditions, while  from
 the Euler  equation   we find
\bea&&
	\frac{\nabla_a p}{p+\rho}=-\nabla_a\ln(U_t)+\frac{\Omega \nabla_a \ell}{1-\Omega\ell},\quad{where}
\\\nonumber
&& \Omega=\frac{U^\phi}{U^t}=-\frac{g_{tt}}{g_{\phi\phi}}\ell_0=\frac{f(r)}{r^2\sin^2\theta}\ell_0,\quad \ell=-\frac{U_\phi}{U_t}=\frac{L}{E}.
\eea
$\Omega$ is the fluid relativistic angular frequency,
 $\ell$  specific angular momenta,  here   assumed  constant  and conserved \emph{per} disk.
Explicitly  we have
\bea&&\nonumber
V_{eff}(\ell)=U_t= \pm\sqrt{\frac{g_{\phi t}^2-g_{tt} g_{\phi \phi}}{g_{\phi \phi}+2 \ell g_{\phi t} +\ell^2g_{tt}}},\quad
\Omega(\ell) \equiv\frac{U^{\phi}}{U^t }=-\frac{{E} g_{\phi t}+g_{tt} L}{{E} g_{\phi \phi}+g_{\phi t} L}= -\frac{g_{t\phi}+g_{tt} \ell}{g_{\phi\phi}+g_{t\phi} \ell},
\\&& \ell=-\frac{U^{\phi} +g_{\phi t} U^t}{g_{tt} U^t +g_{\phi t}U^{\phi}} =-\frac{g_{t\phi}+g_{\phi\phi} \Omega }{g_{tt}+g_{t\phi} \Omega }=
\\
&&\nonumber\ell_{\pm}
=\frac{a^3M +aMr(3r-4M)\pm\sqrt{Mr^3 \left[a^2+(r-2M)r\right]^2}}{[Ma^2-(r-2M)^2r]M},
\eea
where $V_{eff}(\ell)$ is the {torus effective potential} function of the radius $r$ and parameters  $(\ell, a)$ ,  the  function
$W\equiv\ln V_{eff}(\ell)$, is sometimes called Paczynski or   Paczynski-Wiita  (P-W) potential.
 The tori are regulated by  the balance of the    hydrostatic  and   centrifugal  factors due to the fluid  rotation and by the curvature  effects  of the  Kerr background, encoded in the effective potential function $V_{eff}$.
There is
\begin{equation}\label{E:II}
 	\int^p_z\frac{d p}{p +\rho}=W(p)-W(0)=-\ln\frac{U_t}{(U_t)_{in}}+\int^l_{l_in}\frac{\Omega d l\ell}{1-\Omega \ell}.
\end{equation}
 Assuming  the fluid  is   characterized by the  specific  angular momentum  $\ell$  constant (see also discussion  \cite{Lei:2008ui}),  we consider the equation for  $W:\;  \ln(V_{eff})=\rm{c}=\rm{constant}$ or $V_{eff}=K=$constant. We shall describe more extensively the consequences of this procedure on determination on tori morphology  in Sec.\il(\ref{Sec:constr}).
The  function  $V_{eff}(\ell)$   is invariant under the mutual transformation of the parameters
$(a,\ell)\rightarrow(-a,-\ell)$, thus we can limit our analysis to  positive values of $a>0$,
for corotating  $(\ell>0)$ and counterrotating   $(\ell<0)$ fluids.  More generally  we adopt the notation $(\pm)$  for  counterrotating or corotating matter  respectively.

\textbf{From one torus to multi accreting systems}

  In   Sec.\il(\ref{Sec:intro-RAD}) we  comment on the multi-accretions processes resulting in  multi orbiting structures, modelled here in the  ringed accretion disks developed in \citet{pugtot,ringed,dsystem,open,letter}.
This model of orbiting  macrostructure determines the  \emph{set} of tori,  providing   limitations on its existence and stability.
 Ringed accretion disk (\textbf{RAD}) is a fully general relativistic model of
axially symmetric but "knobby"
accretion disk orbiting  a Kerr \textbf{SMBH}.  The specific angular momenta  $\ell$,   assumed  constant  and conserved for each torus,  is a  variable   in the \textbf{\textbf{RAD}} distribution.
We  distinguish a \textbf{RAD} where tori have all possible relative inclination angles (\citet[see][]{mnras3, mnras3a}), considering tilted or warped disks,
 and  the \textbf{eRAD} where all the tori are on the equatorial plane of the central (axially-symmetric or spherically-symmetric) attractor,  coincident with the symmetry plane of each toroid of the aggregate.
To fix the ideas we concentrate here on the \textbf{eRAD}.  The case of \textbf{eRAD} could be seen as a particularization of the  \textbf{RAD} case and, on the other hand,  it has been proved that many results on the equatorial case condition hold or are of immediate extension for the \textbf{RAD} case. We will often refer to a  \textbf{RAD} for the general properties independent of the particularization of the symmetry plane. The \textbf{eRAD}   model  is
  constituted by an aggregate of corotating and counterrotating perfect fluid, (one particle species), tori orbiting on the equatorial plane of  one central  Kerr  \textbf{SMBH} attractor. Each torus is then part of the  coplanar axis-symmetrical structured toroidal disks, orbiting in the equatorial plane of a single central Kerr \textbf{BH},
 the (\textbf{RADs}),  introduced in \citet{pugtot} and  detailed in \citet{ringed,open,dsystem,letter,long}.  For  the  \textbf{eRAD} frame, where all the tori are centered on the \textbf{BH} equatorial plane,  for  the tori couple $(C_{(i)}, C_{(o)})$,    with specific angular momentum $(\ell_{(i)}, \ell_{(o)})$,   we  introduce   the  short notation  $\pp_i<\pp_o$  and  $\pp_o>\pp_i$ for the inner and outer configurations of  the tori couple. This is actually a notation on the relative location of the centers of maximum pressure (and density) in the two disks, which  are fixed by the two effective  potential   functions relative  to the two tori  respectively and by a leading \textbf{RAD} function, representing the distribution of tori    as centers of points of maximum pressure and density in the tori. Therefore we can  introduce   the concept  of
 \emph{$\ell$corotating} disks,  defined by  the condition $\ell_{(i)}\ell_{(o)}>0$, and \emph{$\ell$counterrotating}  disks defined  by the relations   $\ell_{(i)}\ell_{(o)}<0$.  The two $\ell$corotating tori  can be both corotating, $\ell a>0$, or counterrotating,  $\ell a<0$, with respect to the central attractor spin $a>0$.
The model  constrains  the  formation of the centers of maximum pressure and density in an agglomeration of orbital extended matter, emergence of \textbf{RADs} instabilities in the
phases of accretion onto the central attractor and tori collision emergence \citep{dsystem,Multy,letter}.
Alongside the  set of maxima of the density, which are closer to the central \textbf{BH} attractor, there are the minima of  density, which  are instability points. The \textbf{RAD} dynamics is strongly affected by the
dimensionless spin of the central \textbf{BH} and the fluids relative rotation.
Particularly there is evidence of a strict correlation between \textbf{SMBH} spin,
fluid rotation and magnetic fields in \textbf{RADs} formation and evolution\citep{Multy}--we  will consider the magnetic  in Sec.\il(\ref{Sec:influence}).
Eventually the \textbf{RAD} frame investigation constrains specific classes of tori that could be observed
around some  \textbf{SMBHs} identified by their dimensionless spin.
 On  methodological view point, the novelty  of this approach, with respect to the  other  studies of analogue systems in the context of multi orbiting disks, consists primarily in the fact that other   these studies  foresee a  strong numerical effort (often within a  dynamical frame)  with different very specific assumptions on the tori models, for example considering dust,  while  \textbf{RAD} model  analysis focus on pressure supported perfect fluid disks with any barotropic equation of state.  The final configuration can  indeed provide specific initial data on tori configurations as we discuss  constraints on general classes of tori which can be considered for application  in very diversified scenarios,  including  GRMHD setups. Results completely constraint the possible initial configurations with multiple tori considering both the possibility of tori collision and accretion emergence, or  their morphological characteristics.
Together with the leading function there is  the  energy function $K(r)\equiv V_{eff}(\ell(r),r)$, being $\ell(r)$ the leading \textbf{RAD} function as the distribution of angular momentum in the \textbf{RAD},  providing indications on  stability, being  related to the  energetics of \textbf{BH}-accretion disks systems,   and defining relevant quantities as the  mass accretion rate and cusp luminosity.

    The \textbf{RAD}  scenario has   consequences and
ramifications  on several possible phenomena connected to the \textbf{RAD}  structure  and proceeding from having preferred the analytical and global approach. The global and structural aspect of this approach  has to be intended in the sense of  constraints on the \textbf{RAD} as a whole body rather than focus on the details of each gravitating component constituting the orbiting aggregate.
A further interesting argument  for  \textbf{eRAD} and \textbf{RAD}  is the  determination of the relevance of the   disk verticality,  here conditioned by the analysis of the poloidal  projection of the Euler equation.   The  \textbf{eRAD}  is  mainly  a   one-dimensional model for the cluster,  due  to the conditions ensured  by set of  results known as von Zeipel's theorem.  The radial direction  to some extent also fixes the verticality of the ring components   which are geometrically thick disks and the \textbf{RAD} knobby disk.
A more accurate discussion of the verticality of the model is in  Sec.\il(\ref{Sec:influence}) where we discuss the magnetic case.
However it is important to note that  the density in the  each disk  (and clearly in the \textbf{RAD}) is not homogenous. We establish the  maximum and minimum points of density and pressure  in each toroid, and we can relate it to both the $\ell$ and $K$ function.
The introduction of the \textbf{RAD} rotational law coincident  in the HD model  with the distribution $\ell(r,a)$ (or eventually keeping the explicit dependence on the poloidal angle $\ell(r; a,\sigma)$) is a key aspect for the methodological view point, and the construction of the model. This is essentially a distribution of maximum and minimum points of density and pressure in the orbiting  extended matter configuration. The minima  of the (collections of) potentials are seeds for the rings formations, the minima of pressure  closer to the central attractor,  unstable points and they are   the maxima of the potentials. The closer the  unstable  points are to the central attractor and the greater is the centrifugal component  in the force balance equation, leading eventually to proto-jets  open configurations. The closer to the marginally stable circular orbit are the maxima of HD pressure   the lower the centrifugal component. The  tori cetered very far from the attractor have an extremely large centrifugal component  and  no unstable HD mode developing  in a cusp: the quiescence phase of the  torus is supported by largest component of centrifugal force. These models differentiate between $\ell$corotating and $\ell$countterrotating sequences of tori, especially in relation to configurations formed very close or very far form the attractors, and for dimensionless spin of the \textbf{SMBHs} close to the extreme solution ($a\gtrsim  0.97 M$).
 The  role of a leading function includes the   establishment of  the distribution of minima  points.
Although the seeds are  in fact   to grow (in the sense explained in \citet{dsystem,letter})  in geometrically thick disks, the \textbf{eRAD}   models  usually  geometrically thin  disks with a ringed and structured composition, therefore it can be observed having a part of (geometrically) thin model characteristics but    with differential rotating inter disks shells of jets \citep{open,proto-jet}, a distinctive set of internal activities  consistent in inner accretion from  a double  accreting points (from a corotating and a counterrotating torus), tori collisions \citep{letter}, and chaos emergence \citep{Multy}, altogether with establishment of  different kinds of instable processes as runway instability, runaway--runaway  instability \citep{dsystem,letter}, interrupted phases of accretion, presence  of obscuring tori.
  In the case of a \textbf{RAD} around an (almost) static attractor, this can give rise to an embedded  \textbf{SMBH} in a multi-poles orbiting structure where the central  \textbf{BH} horizon  is screened  to an observer at infinity. Eventually this situation   may give rise to collapse of innermost shells in an extremely violent outburst. These limiting configurations of embedded \textbf{BHs} in globuli may be distinguished  by a distinctive path of QPOs emission  and  recognizable by  different constrains of model parameters.
\begin{figure}
\centering
  \includegraphics[width=9cm]{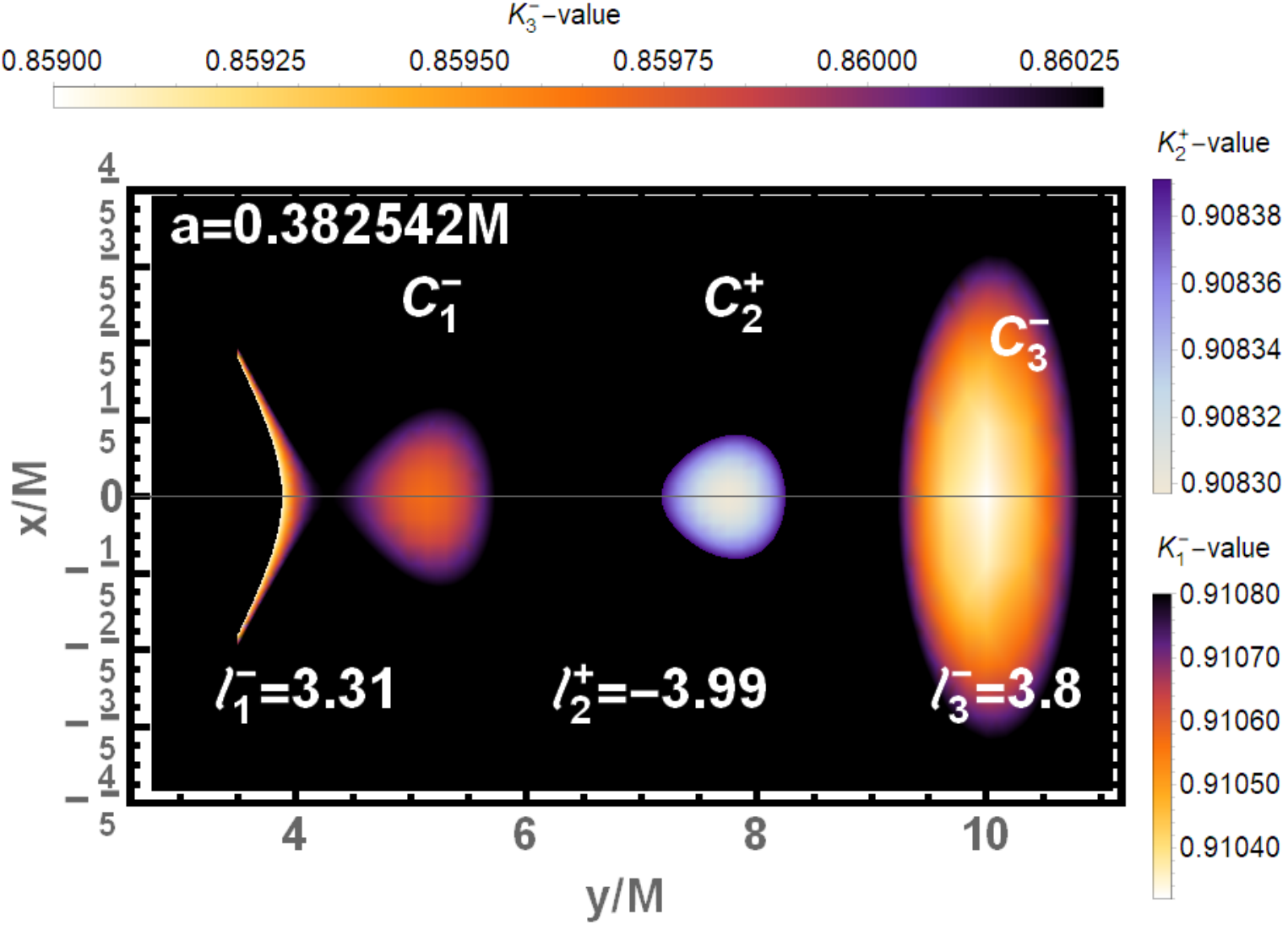}
  \caption{\textbf{eRAD}--Ringed accretion disk of the order $3$, composed by three tori, inner corotating $C_1^-$ with cusp, middle counterrotating $C_2^{+}$ and the outer corotating $C_3^+$ quiescent. The sequence of tori $C_1^-<C_2^+<C_3^-$ is  therefore $\ell$counterrotating. Values of fluid specific angular momentum $\ell^-_1,\ell^+_2, \ell^-_3$ for the three tori are signed on the panel as the \textbf{BH} dimensionless spin $a/M$.  The values of the K parameters, levels of the tori effective potential, $K^-_1, K^+_2, K^-_3$ are in legends.}\label{Fig:developedur}
\end{figure}

\subsection{Constraints, morphology and angular momentum distribution}\label{Sec:constr}
\textbf{The Boyer surfaces}
The described procedure
borrows from the  Boyer theory on the equipressure surfaces for a thick torus \citep{Boy:1965:PCPS:,mnras1}.
Toroidal surfaces correspond to the   equipotential surfaces, with center in the   critical points of $V_{eff}(\ell) $ as function of $r$, thus  solutions of  $V_{eff}=K=$constant.
We can write  essentially the radial gradient  projection $(U^r)$ of the Euler equation, or radial gradient of the effective potentials reducing the problem of accretion disk or \textbf{eRAD} models to  a 1-dimensional problem.
This leads to  four  classes of configurations  corresponding  to  {closed} and  {open} surfaces, and surfaces  with or without
a cusp, i.e. self-crossing open or closed configurations. The
closed, not cusped,   surfaces are  associated to  stationary equilibrium (quiescent)  toroidal configurations.
 For  the  cusped and   closed equipotential surfaces,   the accretion onto the central black hole can occur through the
cusp of the equipotential surface. In this situation  the torus  surface exceeds
the critical equipotential surface (having a cusp), leading to a mechanical
non-equilibrium process where  matter inflows  into the central  black hole (a violation of the hydrostatic equilibrium known as Paczy\'nski mechanism) \citep{abrafra}.
Therefore, in this
accretion model we shortly indicate the  cusp of the  self-crossed closed toroidal surface as  the "inner edge of accreting torus".
Finally, the open equipotential surfaces, which we do not consider explicitly here, have been associated to  the formation of  proto-jets \citep{open,proto-jet}.
We can summarize this situation listing the set of main  resulting configurations,  introducing  also the notation for  \textbf{RAD} tori, as  follows
\begin{itemize}
\item
{${C}$}--cross sections of the{ closed} Boyer surfaces (equilibrium quiescent torus);
\item
{${ C_\times}$ }--cross sections of the {closed cusped}  Boyer surfaces (accreting torus);
\item
{${O_\times}$}--cross sections of the {open cusped}  Boyer surfaces, generally associated to proto-jet configurations. In the following we use the notation $()$ to indicate a configuration which can be closed, $C$,  or open $O$-
\end{itemize}
 Many features of the tori dynamics and morphology as thickness,  stretching in the equatorial plane are predominantly determined by the geometric properties of spacetime via the effective potential\footnote{{This also includes effects related to the disks  energetics such  as  the
mass-flux,  the  enthalpy-flux (evaluating also the temperature parameter),
and  the flux thickness. We
assume     polytropic fluids with pressure $p=\kappa \rho^{1+1/n}$. The mass-flux,  enthalpy-flux and flux thickness  have all  form   $\mathcal{O}(r_\times,r_s,n)=q(n,\kappa)(W_s-W_{\times})^{d(n)}$, where $q(n,\kappa)$ and $d(n)$ are different functions of the polytropic index and constant:  $W_s\geq W_{\times}$ is the value of the equipotential surface, which is taken with respect to the asymptotic value, while $ W_{\times}$ is evaluated at the cusp $r_{\times}$.
Thus there is more specifically
$\mathrm{{{Enthalpy-flux}}}=\mathcal{D}(n,\kappa) (W_s-W)^{n+3/2}$,
$\mathrm{{{Mass-Flux}}}= \mathcal{C}(n,\kappa) (W_s-W)^{n+1/2}$.  The quantity
$\mathcal{L}_{\times}/\mathcal{L}= \mathcal{B}/\mathcal{A} (W_s-W_{\times})/(\eta c^2)$   evaluates   the  fraction of energy produced inside the flow and not radiated through the surface (swallowed by the  \textbf{BH}), $(\mathcal{A}, \mathcal{B},\mathcal{C})$ are  constant depending on the polytropics. The efficiency is
$\eta\equiv \mathcal{L}/\dot{M}c^2$,    $\mathcal{L}$ representing the total luminosity, the $\dot{M}$ is  the total accretion rate where for a stationary flow   $\dot{M}=\dot{M}_\times$  that is  the   mass flow rate through the cusp (mass loss, accretion rates).
Then $\dot{M}_\times$, the cusp luminosity $\mathcal{L}_\times$ (and the accretion efficiency $\eta$),
 measuring the
rate the thermal-energy  is  carried at cusp,
 have  the compact form  $\mathcal{P}=\mathcal{O}(r_\times,r_s,n) r_\times(\Omega_K(r_\times))^{-1}$, where the relativistic frequency $\Omega$  reduces  to the Keplerian  one $\Omega_K$ at the edges of the accretion torus because   the pressure forces   vanish.  There is    $\mathcal{L}_{\times}={\mathcal{B}(n,\kappa) r_{\times} (W_s-W_{\times})^{n+2}}/{\Omega_K(r_{\times})}$, and accretion rate for the disk is   $\dot{m}= \dot{M}/\dot{M}_{Edd}$,
  while  $\dot{M}_{\times}={\mathcal{A}(n,\kappa) r_{\times} (W_s-W_{\times})^{n+1}}/{\Omega_K(r_{\times})}$  \citep{otte0,letter,Multy}. }} $V_{eff}$ (the surface of constant pressure (the Boyer surface) that is orthogonal to the gradient of the effective potential).
The maximum  of the hydrostatic pressure corresponds to the minimum of the  effective potential $V_{eff}$, and it is the torus center $r_{cent}$.   The instability points of the tori, as envisaged by the  P-W mechanics, are located at the minima of the pressure and therefore maximum of $V_{eff}$. To identify these points,  we  therefore need to compute the  critical points of  $V_{eff}(r)$  as function of the radius $r$. Equation  $\partial_r V_{eff}$  can be solved for the specific angular momentum of the fluid  $\ell(r)$.
In fact, the forces balance  condition for the accretion torus  can be encoded in     two functions defining each  \textbf{RAD} component:
the  torus  fluid (critical) specific angular momentum: $\ell^{\pm}(a;r):\;\partial_r V_{eff}=0$,  defining the  critical points  of the hydrostatic pressure in the torus, and the function
$K^{\pm}(a;r,\ell): \; V_{eff}(a;r,\ell^{\pm})$,    for  counterrotating and corotating fluids respectively,  $\ell$ is present as a fundamental feature of the theory of accretion disks.
Curves $K^{\pm}(a;r,\ell)$ locate the tori centers, provide information on torus elongation and  density and,  for a  torus accreting onto the central \textbf{BH}, determine the inner and outer torus edges.
 The  couple of  constant parameters $(\ell,K)$ uniquely identifies each Boyer surface and these can be directly reduced to a single parameter   $\ell$,  in presence of a cusp.
Then, a particularly attractive feature of tori with constant specific
angular momentum $\ell$ is that   the  $\ell$corotating tori and  particularly  $\ell$counterrotating tori  are   constrained   by  the Kerr geometry {geodesic structure} restricting these configurations to the geometric and causal properties of the fixed background.
Boyer surfaces as constant pressure  surfaces as  essentially based here on the application  of von Zeipel theorem.

\textbf{The von Zeipel condition}
In this model the entropy is constant along the flow. According to the von Zeipel condition, the surfaces of constant angular velocity $\Omega$ and of constant specific angular momentum $\ell$ coincide \citep{M.A.Abramowicz,Chakrabarti0,Chakrabarti,Zanotti:2014haa} and  the rotation law $\ell=\ell(\Omega)$ is independent of the equation of state \citep{Lei:2008ui}.
 This aspect of the disk rotational law  is clearly linked to scale-times of the main physical processes involved in the disks, the accretion mechanism for transporting angular momentum in the disk, possibly the MRI process and  the turbulence emergence dependent the magnetic fields and in the vertical structure of the disk.
 All these issues are plagued by a certain degree of freedom. Paczynski realized that as the  assumptions on viscosity involves  eventually an  ad hoc adoption (as the  $\alpha$ prescription), one can assume the distribution $\ell$, which is here settled by   geometric  grounds.
Essentially the application of the so called Boyer condition within the conditions of the  von Zeipel results  reduces to an integrability condition on the Euler equations. In the case of a barotropic fluid, the right  side of  the differential equation  is the gradient of a scalar,  which is possible if and only if it is
$\ell=\ell(\Omega)$.
More specifically this implies  that  if $\Sigma_{\mathbf{Q}}$ is the  surface $\mathbf{Q}=$constant, for any quantity or set of quantities $\mathbf{Q}$, then there is  $\Sigma_{i}=$constant for \(i\in(p,\rho, \ell, \Omega) \),  where the angular frequency  is indeed $\Omega=\Omega(\ell)$ and it holds that  $\Sigma_i=\Sigma_{j}$ for \({i, j}\in(p,\rho, \ell, \Omega) \)\citep{Boy:1965:PCPS:,Raine}.

\textbf{Angular momentum distribution}
An essential part of the \textbf{\textbf{RAD}} analysis is the characterization of the boundary conditions on the toroidal structure  in the orbiting  agglomerate, which constitutes  the  \textbf{RAD} disk inner structure. The model is constructed  investigating the
 function representing the  angular momentum distribution  $\ell(r)$ inside the \textbf{RAD}  disk (which is not constant). This function sets  the toroids  location (and certain equilibrium conditions) in the agglomerate and it coincides, in the hydrodynamical \textbf{RAD} model of perfect fluids,  with the distribution of specific angular momentum of the fluid in each agglomerate toroid (where it is  a constant parameter $\ell$).
In general  in these  models of accretion disks the angular momentum of matter in the  disks  is considered to be sufficiently high   for  the centrifugal force to be  a predominant component of the four forces regulating the disks balance (centrifugal, gravitational, pressure and  magnetic forces, and eventually dissipative effects). In general this holds particularly for  situations where the gravitational background is generated by a \textbf{SMBHs} shaping morphology and a great part of dynamics on (micro-and macroscopical scale of the) disks--as relevance of turbolence/viscosity or  emerging global instabilities and oscillation modes.   The  Bondi quasi-spherical  accretion constitutes an example of situation  when the condition ($|\ell|>|L|$) is not fulfilled. In the  Bondi quasi spherical  accretion, the fluid angular momentum is  everywhere  smaller  than  the  Keplerian  one and
therefore   dynamically  unimportant.
In general    accretion disks, there must be an extended region where there is  $\mp\ell^{\pm}>\mp L^{\pm}$  in the same  orbital region (explicitly  including  counterrotating fluids on Kerr background, it is however possible this condition should be adapted to this special case). This limiting condition is assumed to hold for a general  accretion torus with a general   angular momentum distribution.
However, the models under examinations here are based on   a full GR onset for each \textbf{RAD} toroid, where in fact there exists
an   extended region where the fluids angular momentum in the torus  is larger or equal  (in magnitude) than the Keplerian (test particle) angular momentum. More precisely each toroid  component is a thick, opaque (high optical depth) and super-Eddington, radiation pressure supported  accretion disk (in the toroidal disks, pressure gradients are crucial) cooled by advection with low viscosity.
Consequently, during the evolution of dynamical processes, the functional form of the angular
momentum and entropy distribution depend on the initial conditions of the system and not on
the details of the dissipative processes.
From these considerations, using the distribution of relativistic  specific angular momentum   in the  \textbf{\textbf{RAD}}
in  \citet{ringed,open,dsystem,Multy},   constraints on the range of variation of the inner edge of accreting  torus,   $r_\times$, and on the point of maximum density  (pressure) in each torus, $r_{cent}$, were fixed in dependence from the range of variation of the specific angular momentum in the \textbf{RAD} disk.
It is worth specifying that this strong dependence of the model on the geometric properties  of spacetime induced by the central attractor enables  to  apply to a certain extent the   results for these disks as reliable constraints and reference   to  different models of accretion disks \citet{abrafra}.
(Moreover the  Maxwell stresses, produced by the MRI  driven turbulence,  lead the final distribution of the angular momentum to  approximate the  Keplerian  distribution and  more generally showing independence from  the initial matter distribution--\citep{abrafra}.)
Precisely, constraints  on the ranges of values of the fluid specific angular momentum $\ell$  follow from the geometric background influence which is  summarized in special points on the momentum distribution and composed by the  \emph{marginally stable circular orbit}, $r_{mso}^{\pm}$, the \emph{marginally bounded circular orbit}, $r_{mbo}^{\pm}$ and  the \emph{marginal circular orbit} (photon orbit) $r_{\gamma}^{\pm}$

\begin{figure}
\centering
  \includegraphics[width=8cm]{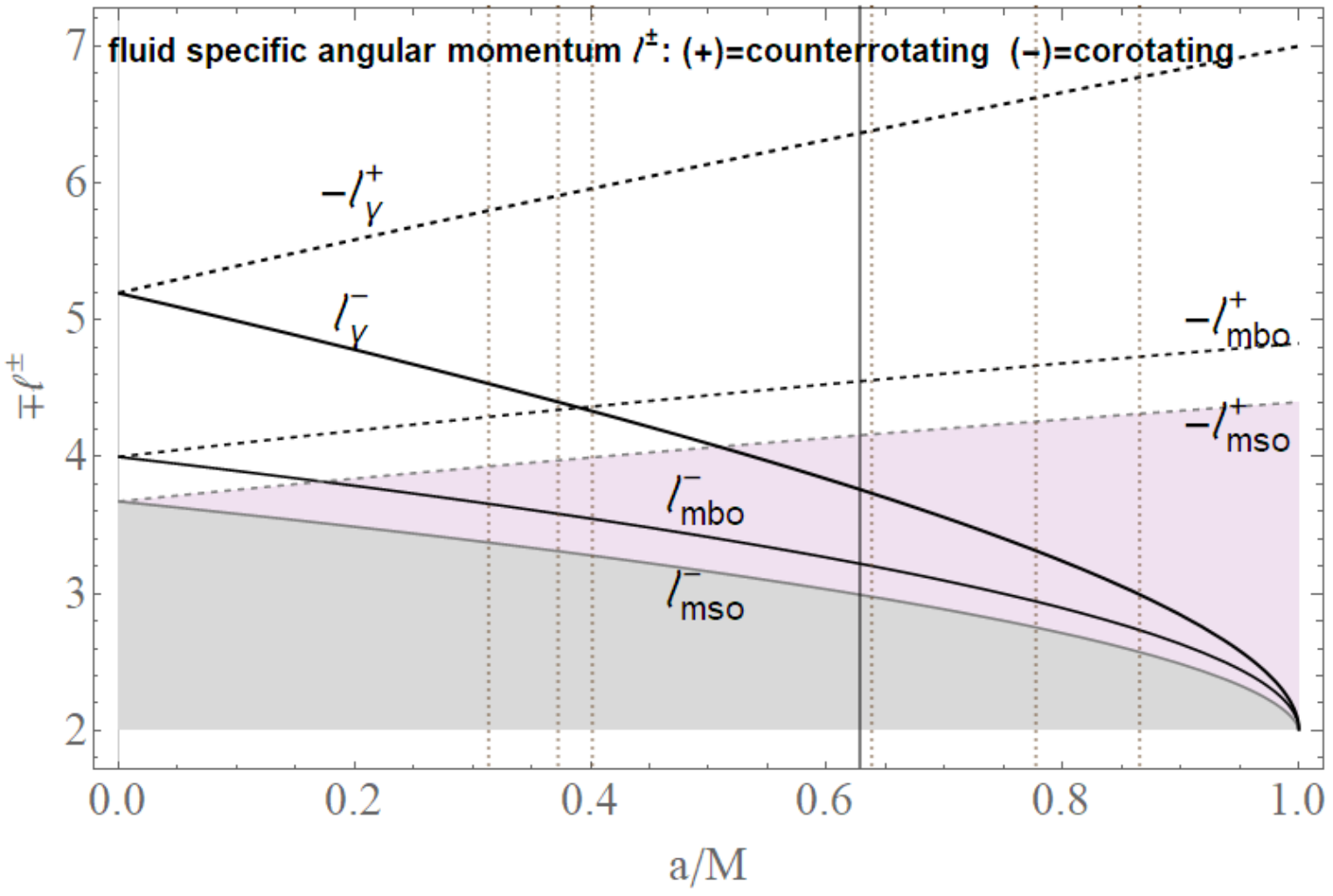}
  \includegraphics[width=8cm]{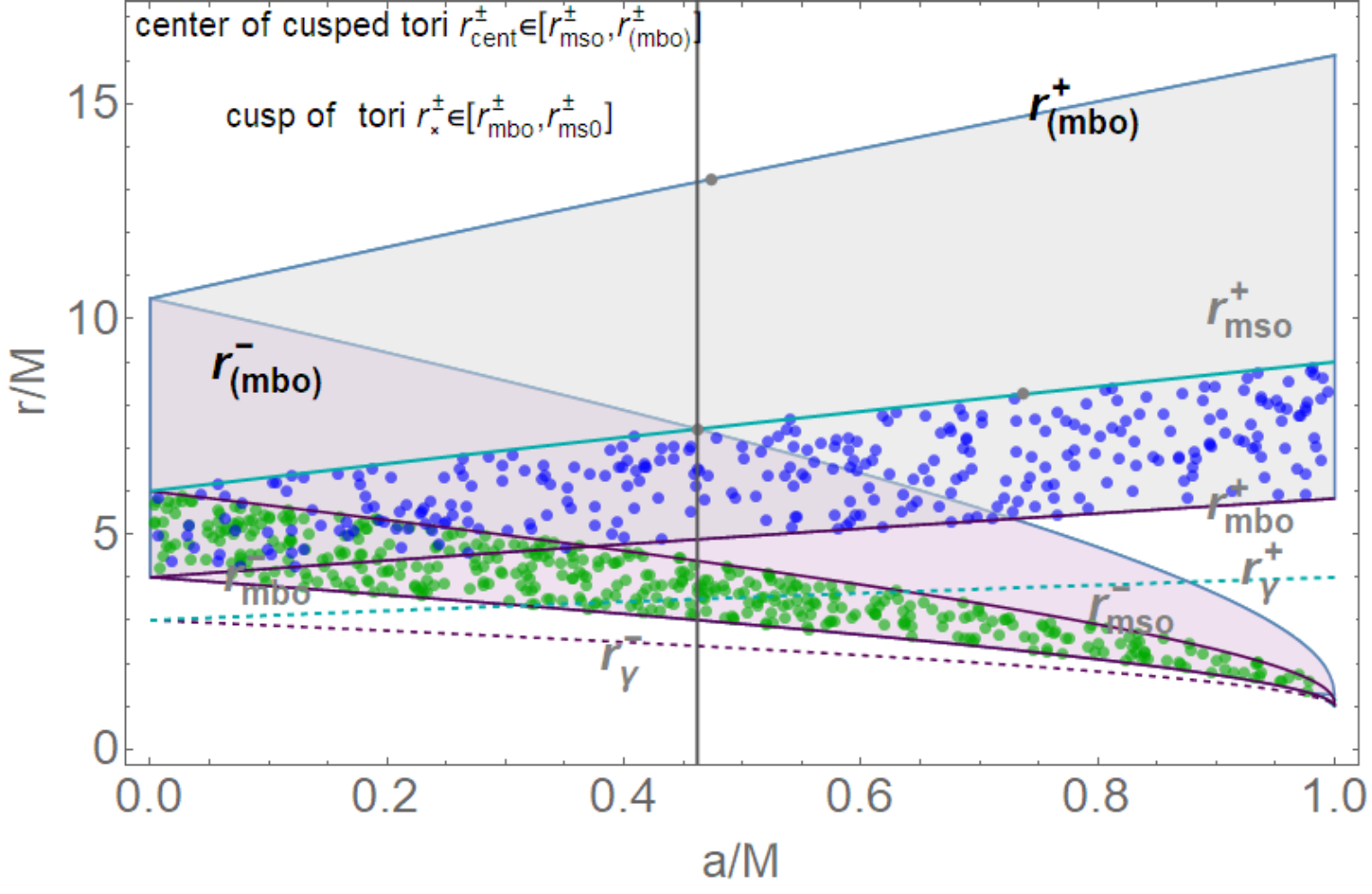}
  \caption{Upper panel: Fluid specific angular momentum $\ell^{\pm}$ of the tori in the \textbf{RAD}, for corotating $(-)$, and counterrotating $(+)$, fluids, versus \textbf{SMBH} dimensionless spin $a/M$. The tori are for $\mp\ell^{\pm}>\mp\ell_{mso}^{\pm}$ respectively. There is   $\ell_{\bullet}\equiv \ell(r_{\bullet})$ for $r_{\bullet}=\{r_{mso},r_{mbo},{r_{\gamma}}\}$, $r_{mso}$  is the marginally stable circular orbit, $r_{mbo}$ is the marginally bounded orbit and $r_{\gamma}$ the last circular (photon) orbit. Bottom panel: radii $r_{mbo}$ and $r_{mso}$ and the pair $r_{(mbo)}$ and $r_{(mso)}$ defined in Eq.\il(\ref{Eq:conveng-defini}) as functions of the dimensionless spin of the \textbf{BH}. Regions of Table\il(\ref{Table:L1L2L3}) are shown. }\label{Fig:PlotxdisollMMb}
\end{figure}

Alongside the geodesic structure of the Kerr spacetime represented by the set of radii $R\equiv (r_{mso}^{\pm}, r_{mbo}^{\pm},r_{{\gamma}}^{\pm})$, and $r_{\mathcal{M}}^{\pm}$  solution  of  $\partial_r^2\ell=0$,
 it is  necessary to  introduce also the   radii $r_{(mbo)}^{\pm}$ and $r_{(\gamma)}^{\pm}$  or more generally $R_{({r})}\equiv (r_{(mbo)}^{\pm}, r_{(\gamma)}^{\pm},r_{(\mathcal{M})}^{\pm})$ {(We include also the radius $r_{(\mathcal{M})}^{\pm}$)}
defined as the solutions of  the  following equations
\bea&&\label{Eq:conveng-defini}
{r}_{{mbo}}^{\pm}:\;\ell_{\pm}(r_{mbo}^{\pm})=
 \ell_{\pm}({r}_{{mbo}}^{\pm})\equiv {\ell_{mbo}^{\pm}},
\quad
  r_{(\gamma)}^{\pm}: \ell_{\pm}(r_{{\gamma}}^{\pm})=
  \ell_{\pm}(r_{(\gamma)}^{\pm})\equiv {\ell_{{\gamma}}^{\pm}},
  \\
  &&\nonumber\mbox{and}\quad
r_{(\mathcal{M})}^{\pm}: \ell_{\pm}(r_{(\mathcal{M})}^{\pm})= \ell_{\mathcal{M}}^{\pm}
\eea
 relevant to the location of the disk center and outer edge  and radius, where
\bea
r_{\gamma}^{\pm}<r_{mbo}^{\pm}<r_{mso}^{\pm}<
 {r}_{(mbo)}^{\pm}<
 r_{(\gamma)}^{\pm}
 \eea
 respectively for the counterrotating $(+)$ and corotating $(-)$ orbits.
This expanded geodesic structure, represented by the union of radii $R$ and $R_{(r)}$ sets, rules large  part of the geometrically  thick disk physics  and   multiple structures in the \textbf{RAD} model. The presence of  these radii stands as one of the main effects of the  presence of a strong curvature  of  the background geometry\citep{Multy,open,long}. In Table\il(\ref{Table:L1L2L3}) we summarize the constraints, defining ranges  of fluids specific angular momentum $(\mathbf{L_1,L_2,L_3})$, see also Figs\il(\ref{Fig:PlotxdisollMMb}).

\begin{table}
\resizebox{.99\textwidth}{!}{%
\begin{tabular}{l}
\textbf{[$\ell\in \mathbf{L_1}$: quiescent  and cusped tori]}--$
\mp \mathbf{L_1}^{\pm}\equiv[\mp \ell_{mso}^{\pm},\mp\ell_{mbo}^{\pm}[$:
\\
    topologies $(C_1, C_{\times})$;  accretion point   $r_{\times}\in]r_{mbo},r_{mso}]$ and center with maximum pressure  $r_{cent}\in]r_{mso},r_{(mbo)}]$;
\\
\textbf{[$\ell\in \mathbf{L_2}$: quiescent  tori and proto-jets]}--$\mp \mathbf{L_2}^{\pm}\equiv[\mp \ell_{mbo}^{\pm},\mp\ell_{\gamma}^{\pm}[ $:
\\
 topologies    $(C_2, O_{\times})$ are possible;   unstable point  $r_{j}\in]r_{\gamma},r_{mbo}]$  and  center with maximum pressure $r_{cent}\in]r_{(mbo)},r_{(\gamma)}]$;
\\
\textbf{[$\ell\in \mathbf{L_3}$: quiescent  tori]} $\equiv\mp \mathbf{L_3}^{\pm}\equiv\ \ell \geq\mp\ell_{\gamma}^{\pm} $:
\\
quiescent  torus  $C_3$  with center $r_{cent}>r_{(\gamma)}$;
\\
 \end{tabular}}
 \caption{It is
 $mso$ for marginally stable orbit and $mbo$ for marginally bounded orbit, $\gamma$ is for marginally circular orbit (photon orbit)--see also Eq.\il(\ref{Eq:conveng-defini}) for details on notation and Figs\il(\ref{Fig:PlotxdisollMMb}) for the representation of the different regions.
}\label{Table:L1L2L3}
\end{table}
These models allow the determination of a wide number  of aspects of disks morphology, dynamics and stability. Although restricted by the typical assumptions of these simplified models, thick (stationary) disks provide a striking good approximation of  several  aspects of accretion instabilities in  different and  more refined dynamical models. Some of these features are the   tori  elongation on their symmetry plane, the  inner edge of quiescent and accreting disks, the tori   thickness, the  maximum height, and the critical pressure points.
It is possible to  evaluate for example  the inner and outer edge of an accretion torus as follows:
\bea\label{Eq:r-in-out-papers-out}
&&
r_{in}(a;\ell,\overline{Q})=-\frac{2}{3}\left[\mathbf{\alpha} \cos\left[\frac{1}{3} (\pi +\arccos\beta)\right]+\frac{1}{\overline{Q}} \right],
\\
&&
r_{out}(a;\ell,\overline{Q})=\frac{2}{3}\left[\alpha\cos\left(\frac{1}{3} \arccos\beta \right)-\frac{1}{ \overline{Q}}\right],
\\
&&\nonumber
\mbox{where }
\quad
\alpha=\sqrt{\frac{4+3 \overline{Q} \left[(\overline{Q}+1)(\ell^2-a^2) +a^2\right]}{\overline{Q}^2}},\quad\overline{Q}\equiv(K^2-1)<0, \\
&& \beta =-\frac{9 \overline{Q} \left[(\overline{Q}+1)(\ell^2-a^2) +a^2\right]+8+27(\overline{Q}+1) \overline{Q}^2 (a-\ell)^2}{\alpha^3\overline{Q}^3}
  \eea
(dimensionless units have been used).
The use of notation \emph{\textbf{(i)}}: $\pp_i^{\pm}<\pp_o^{\pm}$ and  \emph{\textbf{(ii)}}: $\pp_i^{\pm}<\pp_o^{\mp}$,   for a tori couple in a \textbf{eRAD}  highlights some properties of these structures. Using the  couple seeds  for the  two \textbf{eRAD} configurations in notation \emph{\textbf{ (i)}} for a $\ell$corotating and \emph{\textbf{(ii)}} and a $\ell$counterotating couple, one can characterize different phases of possible \textbf{eRAD} evolution.
 Also, symbols $\lessgtr$ $\left(\lessgtr_{\mathbf{\times}}\right)$
 for two tori refer to the relative position of the {tori centers}   $r_{cent}$ (accretion points or unstable points $r_{\times})$:
thus, for example, we use short notation $(\pp^-<\pp^+,\pp^-<_{\mathbf{\times}}\pp^+)\equiv \pp^{-}\ll_{\mathbf{\times}}\pp^+$  that means $r_{cent}^-<r_{cent}^+$  and  $r_{\times}^-<r_{\times}^+$.
In the \textbf{eRAD} the seed couple introduced in \citet{dsystem} is in fact a pair of tori, intended as constructive  seed as it serves from the methodological  viewpoint to manage the constraints on the effective potential of the entire \textbf{eRAD} structure introduced  in \citet{ringed} and more generally to study the entire structure as a whole orbiting body with a complex and diversified intern. Conversely, the introduction of a law of rotation as the leading function for \textbf{eRAD} and \textbf{RAD} is intended as the distribution of the points of maximum pressure and minimum pressure in a generic distribution of extended orbiting matter with some symmetries - it should then be understood that it  must be specified  how to consider  properly an extended matter  structure in the sense of this  rotational law. Each point in the stability area  of the   leading function, that is to the right   $r>r_{mso}^{\pm}$of the discriminant radius $r_{mso}^{\pm}$ (a minimum of $\pm\ell^{\mp}(r)$),  respectively for the two directions of rotation, has to be  understood as  pressure/density seeds  for  toroidal rings growing. The  regions for corotating and counterrotating fluids  are superimposed depending on the spin value $a/M$. According to different ranges of $a/M$ values, there are several aggregation structures  seeds of super-Keplerian matter in the sense of $|\ell|\geq |L|$, this makes the analysis of the \textbf{eRAD} complex but also provides a surprisingly refined   method to obtain  restricted constraints for the  \textbf{RAD}-\textbf{SMBH} systems.  {(We specify that here as elsewhere the term Keplerian  refers in short  to the motion of a free particle located on  circular orbit around.). } This is mainly due to the fact that orbital regions with boundary defined by the intersection of the radial ranges from the  $R$ and $R_{(r)}$ sets with the additional couple $(r_{\mathcal{M}},r_{(\mathcal{M})})$  strongly depend on the dimensionless spin  of the attractor especially for the $\ell$corotating couples, with an even more complex structure for the $\ell$corotating tori sequences composite  of corotating tori. Furthermore there is the addition  complication,  for the orbiting multi-structures, due to the fact that corotating and counterrotating structures must be considered together, substantially coupling the two separated analysis with further combinations of finite but complex constraints-- \cite{dsystem,letter,Multy}. We show an aspect of this condition below by providing an example of the limits imposed on the attractor spin. In this context, the radii $r_{(\mathcal{M})}$ in the unstable region and  $r_{\mathcal{M}}>r_{mso}$ in the stability region,  must also be interpreted  as extreme  in the distribution of seeds of tori (the stability points) and instability points (obviously corresponding to the left range $r<r_{mso}^{\pm}$). These are interpreted as the  maximum  of distributions of  points of the aggregation seeds in accordance with the maximum and minimum points  of pressure. It can be shown that these radii are related to the derivatives of certain frequencies of oscillations typical of  thick toroidal structures \citep{Fi-Ringed}.
According to different  conditions  set on the \textbf{RAD}, there are further  limits to the inequality $|L|<|\ell|$.
Accordingly, for example, there are the following critical values of  the spin:
\bea\nonumber
&&a_{1}\equiv0.4740M:\quad r_{(mbo)}^+=r_{({{\gamma}})}^-,
\quad
 a_{2}=0.461854M:\;r_{(mbo)}^-=r_{mso}^+
 \\
 &&
  a_{3}\equiv0.73688 M:\;r_{({\gamma})}^-=r_{\mso}^+,
\eea
These identify, on the basis of the doubled geodesic structures classes of attractors where specific \textbf{eRAD} can be observed \citep{Multy}.
\subsection{Multi-accretion processes, multi--orbiting structures  and ringed accretion disks}\label{Sec:intro-RAD}
 The \textbf{RAD} structures pose the interesting issue  of the internal activity of the  tori ringed cluster and particularly the internal exchanges of energy and matter between the tori and the tori and central \textbf{SMBH}.  In this respect, \textbf{eRAD } system, \textbf{RAD} and globuli are characterized by a vivid internal activity made up of collision among tori, inner  accretion, internal jet launching, whether constrained  by open-proto-jet configurations considered here,  in conjunction with an inner edge-accretion jet emission correlation. These structures  can also constitute a basis for interpretation of   the mass accretion rates of supermassive black holes in \textbf{AGNs}, and several other phenomena connected with energetics of the accretion disks, but also the evolution  of the central attractor with particular respect to its spin evolutions.
The \textbf{RADs} model follows the possibility that more  accretion orbiting configurations  can form  around very compact objects  in the special environment of the \textbf{AGNs}-\textbf{BHs} and Quasars. Arising from different \textbf{BHs} accretion periods  and from the   host Galaxy life,  such configurations can report, in their characteristics, traces  of the different periods   during several accretion regimes occurred in the lifetime  typical of non-isolated  Kerr \textbf{BHs} {\citep{Aligetal(2013),Blanchard:2017zfe,long,NixonKing(2012b)}.
During  the  evolution of   \textbf{BHs} in these environments both  corotating  and counterrotating accretion stages are
mixed   during various accretion periods  of the attractor  life \citep{Lovelace:1996kx,Carmona-Loaiza:2015fqa,Dyda:2014pia,Volonteri:2002vz},  thus \textbf{RADs} tori  may be even  misaligned \citep{Aly:2015vqa}.
An example of \textbf{eRAD} composed by three tori  is in Fig\il(\ref{Fig:developedur}).
These systems, including accretion
disks misaligned with respect to the spin of a central black hole, formed  in \textbf{AGNs}, may be consequential to periods of chaotical accretion,
where  corotating and counterrotating tori and strongly misaligned disks  can be formed.

Misaligned disks  have been studied  for example in \citet{King:2018mgw,Aly:2015vqa,NixonKing(2012b),Kovar11,Slany:2013rml,Kovar:2014tla,Kovar:2016kqh,Schroven:2018agz}.
 However, in the context of the misaligned tori, a  Kerr black  hole plays a relevant  role both  in the disk structure and for the effects that the disk  evolution has  on the evolution of the  central \textbf{BH} spin.
In the context of the misaligned tori, a \textbf{BH} warped torus evolves together with its attractor changing its mass, the magnitude of its spin, and the spin orientation.  The impact of the \textbf{BHs} spin on the misaligned disks  merges with the complexity of a \textbf{RAD} structure reflecting in the Bardeen-Petterson effect, causing shift of the tilted accretion disk into the \textbf{BH} equatorial plane  due to the combined effect of the disk inclination and the frame dragging of the Kerr spacetime \citep{BP}.
 \textbf{RAD} structures are governed mainly by geometry of the Kerr \textbf{SMBH} attractors-- \cite{ringed,open,dsystem,proto-jet,long,letter,mnras1,Fi-Ringed,Multy}.
In the \textbf{eRAD} a  tori couple (a state) with  only the inner accreting (cusped) torus  and an outer accreting counterrotating  configuration  is possible, keeping the   stability of the \textbf{RAD}  structure. The misalignment  of the toroidal structures  allows to reconsider in some extent the possibility of the presence of multi accreting  tori on different planes, enlightening interesting situations and phenomenologies   which were   not allowed for the \textbf{eRAD}. There is the possibility that   the presence of more accreting \textbf{RAD} tori  (including collisional regions) could increase the accretion rate of the central \textbf{BH}.

 In the \textbf{RAD} composed by misaligned tori  it has been explored the hypothesis of a quasi-complete covering of the \textbf{BH} horizon with a \textbf{BH} embedded in a accreting ball made by a composition of \textbf{RAD} tori  having  different orientations of the fluids specific angular momentum (tori spins). In the \textbf{RAD}  context,   under  special conditions  (depending on number of toroidal components and tori geometrical thickness),   the system of  \textbf{BH}  entangled tori  could be considered as a  matter embedding,  covering the central \textbf{BH} from a distant observer at different angles.  A globus of clustered  tori is constituted by  a multipoles orbiting matter embedding surrounding the central \textbf{BH} converging  the \textbf{BH} horizon.
These   globuli, being  constituted  by orbiting matter   with different  spin orientations constitute in this sense  a multipole   orbiting configuration with a central  \textbf{BH} object that may have significant role in different epochs of the \textbf{BH} life.

\textbf{RAD phenomenology}

\textbf{RADs} Instabilities may reveal therefore  of  significance for the High--Energy Astrophysics related especially
to accretion onto supermassive \textbf{BHs}, and the extremely
energetic phenomena occurring in Quasars and \textbf{AGNs}
that could be observable by  the X-ray observatory \texttt{\textbf{ATHENA}}\footnote{\url{http://the-athena-x-ray-observatory.eu/}}.
From the phenomenological viewpoint, the \textbf{RAD}  model constitutes a  shift in paradigma  from the interpretative framework of the \textbf{BH}-disk interaction  to the \textbf{BH}-\textbf{RAD}  and  clearly opens a broad scenario of investigation.   The \textbf{RAD} model opens the possibility to  review  the main template of  analysis from a \textbf{SMBH}-disk  framework to a \textbf{SMBH}-\textbf{RAD} one.
 The  \textbf{eRAD}  is constructed as   a whole, geometrically thin disk. The current interpretative framework  of the \textbf{BH}-accretion disk physics
 generally foresees the scenario of a \textbf{BH}-one disk system, implying  therefore the possibility  of a ``\textbf{RAD} in disguise'', i.e.   a \textbf{RAD}  could be observed  as a geometrically  thin,  axis-symmetric disk, centered  on the equatorial plane on a  Kerr \textbf{SMBH},  with a ``knobby'' surface and     characterized by   a differential  rotation    with   peculiar optical properties, particularly the  X-ray emission is  expected to show the ringed  structure  in a discrete emission profile.
 In this respect we can consider the analysis in  \citet{S11etal,KS10,Schee:2008fc,Schee:2013bya}  where  relatively indistinct excesses
 of the relativistically broadened  emission-line components were predicted arising in a well-confined
radial distance in the accretion structure, supposed to be
  originated by a series of  episodic accretion events. More generally the  \textbf{AGN} X--ray variability suggests connection between X-rays and the innermost regions of accretion disk. This makes reasonable to  reinterpret  the observations  analyzed so far in the single-torus framework,    in a new interpretive frame represented by the possibility of a multi-tori system.
\textbf{RAD}  are characterized by  peculiar phenomena associated to the instabilities, as the occurrence of double accretion and  its   after-dynamics, the  inter disks proto-jet emission and  the screening tori.
Instabilities should be treated in accordance with a global point of view where the  macrostructure  is considered  as a single, unique   disk  orbiting around  a central \textbf{SMBH}.
 For example  the
radially oscillating tori of the couple could be related to
the high-frequency quasi periodic oscillations
observed in non-thermal X-ray emission from compact
objects. These emissions  could be interpreted as  fingerprints of the characteristic  discrete radial profile of the ringed  structure.

\textbf{Notes on the   RAD instabilities}
We conclude this section with further   comments on the stability of the \textbf{RADs}  configurations.
 It is clear  that the instability of each \textbf{RAD} component  must reflect in an  inter-\textbf{RAD} disk activity.
More in general, there are different \textbf{RAD} instabilities.  A destabilization of the system structure may arise after the  emergence of  an instable phase  of one component of the \textbf{RAD}, for example after an accretion phase of one torus onto the central  \textbf{BH} or  the proto-jet emission  which  is capable to destabilize the entire disk \cite{long,proto-jet}. This case however has been strongly constrained.  The maximum number of accreting tori  in a \textbf{eRAD} is $n_{\times}=2$,  occurring for the couple $\cc_{\times}^-<\cc_{\times}^+$, made by  an inner corotating   and an outer counterrotating  torus  accreting  on the   \textbf{BH}. Secondly a \textbf{RAD} can be destabilized after  collision of a pair of  quiescent tori  of the agglomeration. Collision  may arise for example  after  growing of one torus \citep{dsystem,letter}. In  the couple  $\pp^-<\cc_{\times}^+$, the  accretion phase of the  outer torus   and the collision emergence  can  combine  establishing a complex phase of \textbf{RAD} destabilization.
The particular case of the  emergence of collision for two \textbf{RAD} tori was considered in    \citet{dsystem}, while
interacting  tori and energetic of associated to these processes
     were investigated in  \citet{letter}.   The mass accretion rates,  the  luminosity at the cusps and other fundamental characteristics of the \textbf{BHs} accretion disk physics were also evaluated.
In  \citet{ringed},  \textbf{RAD} perturbative approaches  have also  been described, in  \citet{open},
the emergence of unstable tori have been detailed,  while further discussion on \textbf{RADs} as remnants of  \textbf{AGN} accretion periods are in  \citet{long}.

A further interesting case  is the sequence of three configurations  $\pp^-<\cc_{\times}^+<\cc^{\pm}$  which is a system  where the inner, accreting or quiescent, torus
 can be an obscuring inner torus. Matter, from the outer counterrotating torus, impacts on the corotating inner one, which is screening the accretion  from the central \textbf{SMBH}.
 Collision however may give rise to several possibilities. A possibility consists in the formation of  in a single torus,  in fact canceling the \textbf{RAD} structure. This is a phenomenon which may affect  mainly  the first evolution phases of the  formation of the aggregate. An inner torus of the orbiting \textbf{RAD} couple may form as axially symmetric corotating toroidal disk after a first phase of formation of the outer aggregate component.
Another possibility is the occurrence of a ``drying-feeding'' phase,  involving  interrupted stages of   accretion of one or two tori of a couple:
matter flows between the  two tori of the couple give raise to a series of  interrupted stages of accretion onto the central \textbf{SMBH}. This particular effect,  considered in \citet{dsystem,long} and detailed  in \citet{letter} can  model  the  different phases of super-Eddington accretion  advocated as a mechanism to explain  the large masses observed in \textbf{SMBHs} at high redshift--see for example  \citet{apite1,apite2,apite3,Li:2012ts,Oka2017,Kawa,Allen:2006mh}.
It is however  necessary to consider the oscillations and instabilities associated with the each  component of the aggregate--eventually  this can be also related  to  \textbf{QPOs} emission--see \citet{Montero:2007tc}.
More precisely, the collision instability can lead to different evolutive  paths  for the aggregate tori, depending on the initial conditions of the processes as  the torus rotation with respect to the black hole,  the range of variation of the mass of the torus  and of the  magnitude of the  specific angular momentum of the fluids.
We also mention  the  case of the runaway-runaway instability which is a combination of runway instability and collision of the inner disk with the outer configuration  or, as analyzed in \citet{Multy}, the combination of chaotic effects induced in the  accretion rate.
 A further important features of thick disks is   the Papaloizou-Pringle (\textbf{PP})  global incompressible modes  and the  Papaloizou-Pringle Instability
(\textbf{PPI}),
which is the  global, hydrodynamic, non-axis-symmetric instability   which can combine eventually with parallel  emerging Magneto-Rotational Instability   modes.
In the  geometrically \textbf{HD} thick disks, the accretion process   is strictly interwoven  with the development of the \textbf{PP} instability:  the  mass  loss in  the   Roche  lobe  overflow  regulates  the  accretion  rate  in  the  innermost  part  of   torus. This   self-regulated process on one side
locally  stabilizes   the accreting torus from the   thermal  and  viscous   instabilities    and, on the other side, it   globally  stabilizes the torus from the \textbf{PPI}.
As mentioned above, in these disks,  the \textbf{PPI} hydrodynamic instability  can  entangle with an  emerging  MRI which triggers  eventually predominant larger modes of oscillation (smaller length scales) with respect to typical \textbf{PPI} modes, and  creating a far richer and complex scenarios for  the torus equilibrium properties.
(Note also  that the amount of overflow    may  be also modulated by global disks oscillations).  On the other hand, the
Papaloizou-Pringle (global and non-axis-symmetric) Instability
 is able to transport angular momentum outwardly in the disk and therefore   able to finally trigger the  accretion
\footnote{The  global non-axis-symmetric hydrodynamic (\textbf{HD}) \textbf{PPI}
 implies also   the  formation of long-lasting,
large-scale structures that may be  also  tracer for such tori    in the
in the gravitational wave emission--see for example \citep{f-x-Kiuchi}.}.
In this respect we should note that global instabilities  are affected by  the
  boundary
conditions assumed for the system. In the case of  \textbf{PPI} in  \textbf{RAD} accreting \textbf{HD} tori, for which  the disk inner and outer edges  are well defined,   the \textbf{PPI} is generally suppressed, stabilizing the disks by the accretion flow driven
by the  pressure forces across the cusp, $r_{\times}$.
 The relative importance of  MRI and \textbf{PPI}  and  the interaction of two  processes depends in fact  on many factors and conditions.
In particular in the \textbf{RAD} scenario different factors can be determinant:
 the (turbolent) resistivity, the  emerging of a  dynamo effect, the study  for counterrotating  (retrograde) tori, the  disk self-gravity,  the gravitational interaction between the disk and the central Kerr \textbf{SMBH}  and   the runaway instability  are further aspects which may  contribute  importantly to the characterization of the ongoing  processes\footnote{For example in \citet{Bugli}, using three-dimensional GR-MHD simulations  it is   studied the
interaction between the \textbf{PPI} and the MRI considering  an analytical magnetized equilibrium solution as initial condition. In the \textbf{HD} tori, the
\textbf{PPI} selects the large-scale $m = 1$ azimuthal mode as the fastest growing and non-linearly dominant mode. In different works it is practically shown that  even a
weak toroidal magnetic field can lead to  MRI development which leads to   the suppression of the large-scale modes.
 Notice also that the   magneto-rotational instability  in the disks is  important because disks  can be   locally \textbf{HD} stable (according to  Rayleigh criterion), but they  are unstable for  \textbf{MHD} local  instability  which is
linear and
independent by the  field strength and orientation, and  growing up
   on dynamical time
scales. The torus  (flow) is \textbf{MHD} turbulent due to the MRI.}.
{In Sec.\il(\ref{Sec:influence}), we investigate the
magnetized tori endowed with a   toroidal magnetic field, grounded on the HD models investigated here.}

\section{Magnetized tori:  toroidal magnetic field in multi-accreting tori}\label{Sec:influence}
In \citet{epl,Fi-Ringed} \textbf{RADs} framework has been used  to investigate the influence of the magnetic field  in the formation of the  torus, as a limiting  case  of the \textbf{eRADs} and hence in the formation of the multiple accreting events occurring around a Schwarzschild and Kerr \textbf{SMBH}. Differences between the magnetize case and the HD models, are particularly evident    in the unstable  phases due to the tori collision  and the accretion.
In this section we consider \textbf{RAD} with toroidal sub-structures regulated by the presence, in the force balance equation, of  a toroidal magnetic field component. We refer to the analysis of \citet{Fi-Ringed,epl}. The toroidal magnetic field form used here is the well known Komissarov-solution \citep{Komissarov:2006nz}, used in the approach \citep{epl,Fi-Ringed,proceeding}, see also \citet{abrafra,Luci,epl,adamek,Hamersky:2013cza,Karas:2014rka,
Slany:2013rml,Kovar11,Fragile:2017lbx,Gimeno-Soler:2017qmt}.
This is a  well known solution of toroidal magnetic field widely adopted  in the  axis-symmetric  accretion configurations.
The  magnetized torus of \textbf{RAD} is   widely used to fix up the initial configurations for  numerical integration of a broad
variety  of  {GR-MHD} models.
In our application  here we  generalize this approach to the    discussion of  thick   disks and their morphology reflected by the presence of magnetic field and secondly  the analysis of angular momentum distribution in the aggregate context of \textbf{eRAD}. More precisely in  this section we consider axially symmetric spacetimes  as   Kerr  or the static limit of the Schwarzschild exact solutions.  We also mention  \citet{Zanotti:2014haa} for a discussion on the case of a poloidal magnetic field where a  metric representation  is  adapted to the direction of the field,  as proved by  \citet{10,11}. This work comments on the application and role of von Zeipel theorem for this particular case of introduction of poloidal magnetic field.
The choice of a purely azimuthal (toroidal)  magnetic field is  particularly adapted to the  disks symmetries considered here  and largely adopted  as initial setup for numerical  simulations  in  several general relativistic magnetohydrodynamic  models sharing similar symmetries with the \textbf{RAD} and thick disks  considered here\citep{Luci}.

Magnetic field can have a major influence in the \textbf{BH}-accretion disk systems, especially during the early stage of   tori  formation and the  final steps of evolutions towards the accretion onto the spinning  \textbf{BH},  a phase where  predominant   instabilities occur for  the accreting torus as well as for the \textbf{RAD} system.
The tori rotational law (specific angular momentum) is therefore expected to depend on the  magnetic field.
The counterrotating  and $\ell$counterrotating cases  are proved   significantly that  the  toroidal magnetic field plays an essential role in determining  the disk structure and stability, showing that also a purely azimuthal field is capable to discriminate the  \textbf{eRAD} features.

One question faced in \citet{epl} is if  thickness of a toroidal accretion disk  spinning around a Schwarzschild or Kerr  Black hole can be effected under the influence of  a toroidal magnetic field and by varying the fluid  angular momentum. In  the   HD  case, where there is only a centrifugal component  originated by the fluid rotation  or eventually the  dragging of the central attractor, the  torus thickness remains basically unaffected but tends to increase or decrease slightly  depending on the  gravitational and centrifugal effects   which the disk is subjected to.
In \citet{epl}    it is addressed  for example   the specific question of torus squeezing on the equatorial planes of the central attractor, exploring the disk thickness changing  the physical characteristics of the torus  as the angular momentum and the effective potential  which characterize the  toroidal surface. From the methodological viewpoint, the magnetic field contribution has been  then  considered as part of the  exact GR effective potential  functions.
This analysis  focuses on  role that a toroidal field could have in determining  the  morphology when considered as part of the pressure and modification of effective potential  according to the model developed in Sec.\il(\ref{Sec:Stationar}).  Moreover a consequent problem consists in questioning the role  that a  toroidal magnetic field could play as  a contributing factor to the collimation or launch  of a jet matter component  along  disk  and \textbf{BH}  rotation axis ---\citep{proto-jet}.

The second issue addressed in this section is  a brief discussion on the angular momentum distribution and part of the stability analysis   in the \textbf{eRAD} agglomerate.
 In \citet{Fi-Ringed} it is    analyzed the  effects of a toroidal magnetic field in the formation of  several magnetized   accretion tori  orbiting around one central Kerr \textbf{SMBH}  in \textbf{AGNs}, where
both corotating and counterotating disks are considered on its equatorial plane, while the analysis of \textbf{RAD} misaligned in globular form around   a static attractor is developed in \citet{mnras3}.
This analysis leads therefore to constrain  tori formation and emergence of \textbf{RADs} instabilities,   accretion  onto  the central attractor and   tori collision emergence, are investigated.
Concerning the instability it is convenient to remind here that geometrically thick  disks are subjected to several oscillation modes.  One  modes set  is  constituted by incompressible and  axis-symmetric  modes    corresponding  to  global  oscillations   for radial,  vertical  and   epicyclic  frequencies together with surface  gravity,
acoustic   and  internal    modes  which are  recovered from  the  so called  relativistic  Papaloizou-Pringle (\textbf{PP})
equation--\citep{abrafra}.
The introduction of a purely toroidal and   even  small magnetic field  (considering  the  magnetic pressure \emph{versus}  gas pressure as  defined by the $\beta$-parameter) can have influence on the development of these modes. This is relevant particular for the   torus global  non-axis-symmetric modes,  because of   the generation of the  Magneto-Rotational Instability  due to the  magnetic field and  fluid differential rotation.
The presence of a magnetic field contribution in the disk force balance leads  to a more complex situation where the \textbf{PPI} has to be considered  in a broader context. More generally, whether or  not the hydrodynamical  oscillation  modes   in MHD geometrically thick disks may  survive  such  global  instabilities  or  the  presence  of  a  weak  magnetic field  would strongly affects these, is still under  investigation.
The  linear development of the \textbf{PPI} can however be  affected by the
presence of a magnetic field and by a combined growth of  the MRI.
 These two processes can coexist, enter into competition or  combine depending on local parameters of the model (strongness of the magnetic field as evaluated by $\beta$ parameter). Some studies  seem to suggest that under certain conditions on the strength of the magnetic field and other conditions on the torus onset, this situation can also be resolved in the \textbf{PPI} suppression by the MRI in  the relativistic
accretion disks.

In the case of
  geometrically thick torus endowed with a (purely) toroidal  magnetic field,  considered  here with  the analytic  Komissarov solution, a series of recent analysis shows that  torus is violently prone to  develop the non-axisymmetric MRI in 3D which could  disturb this configuration on dynamical timescales--see \citep{Del-Zanna,Wielgus,Das:2017zkl} and \citep{Bugli}.
 In  the magnetized tori,  as the \textbf{RAD} tori considered in this section, the accretion is
triggered at much earlier times  then in the \textbf{HD} tori, and modes higher then the azimuthal  $m=1$ mode, typical of  \textbf{HD-PPI} tori, emerge together with $m=1$.  GR-MHD investigations  show  generally an increase of   turbulent kinetic energy in the   earlier phases competing with the GR-HD ones, consequently  accretion   is in fact  triggered by the Maxwell     stresses instead of the  \textbf{PPI}.
Eventually  the fundamental
mechanism responsible for the onset of the \textbf{PPI} does not appear to be the predominant one or even to arise at all in the MHD torus. In the magnetized case  on the other hand there is a  broader range of excited frequencies with respect to the GR-HD model.
In general   recent analysis  show that the  inclusion of a toroidal magnetic field could strongly  affect, even with a
sub-thermal  magnetic field, the  \textbf{PPI} (there are suggestions that the action of MRI suppresses the \textbf{PPI}  $m=1$ mode growth).
This may have a relevant  consequence in the double \textbf{RAD} system as MRI stabilizes the disks
to \textbf{PPI}  with  \textbf{MHD} turbulence.
The evaluation of the accretion rates in  the GR-HD double \textbf{RAD} systems has been carried out in \citet{letter}.
It appears however  that  MRI  is more effective and  faster in transport of  angular momentum across
the disk, and higher  accretion rates were proved to occur
in the magnetized models.
Finally it should be noted that,  according to \citet{Fragile:2017lbx}, strong  toroidal  magnetic fields are rapidly suppressed  in this tori, in favor of weaker fields (decrease of $\beta$ parameter).

%

\medskip

In the following we
first  set  the  problem writing the main equations  and discussing  leading function for the  tori distributions in the \textbf{eRAD} agglomeration. Then considerations on magnetic fields and  the choice of the magnetic  parameters values follow. The section closes with
 some  considerations on aspects
of the magnetize disks morphology and  stability. In this section we use mainly dimensionless units.

\medskip

\textbf{Set-up review}

\medskip

As noted in \citet{Komissarov:2006nz} the presence of a magnetic field with a relevant toroidal component can be   related  to the disk differential rotation, viewed as a generating  mechanism of the magnetic field,  for further discussion we refer  to \citet{Komissarov:2006nz,Montero:2007tc,Parker:1955zz,Parker:1970xv,Y.I.I2003,R-ReS1999}, while we refer to   \citet{epl,adamek,Hamersky:2013cza,Karas:2014rka,abrafra} where this solution is dealt  in detail in the context of accretion disks.
In the magnetized case,  following \citep{epl,Fi-Ringed},  therefore we consider  an infinitely conductive plasma where $F_{ab}U^a=0$,    where  $U^a B_a=0$,  with    $\partial_{\phi}B^a=0$ and $B^r=B^{\theta}=0$.
We obtain the relation  $B^t=U^{\phi} B^{\phi}/U^t=\Omega  B^{\phi} $
\footnote{We note that,  with particular symmetries of the background (static) if we set $B^r=0$ from the Maxwell equations, we infer
\(
B^{\theta}\cot\theta=0
\)
(with $\ell=$constant), that is satisfied for $B^{\theta}=0$ or $\theta=\pi/2$.}.
We can write the Euler equation   (\ref{Eq:preg-t}) as
\bea\label{due}
\frac{\partial_*p}{\rho+p}&=&G_*^{(f)}+G_*^{(em)},\quad
%
G_{*}^{\natural}=-\frac{\partial}{\partial*}W^{\natural}_{*} \quad \mbox{where} \quad W_{*}^{f}\equiv\ln V_{eff}, ,
\\\label{Eq:wgg}
&&
W_{*}^{(em)}\equiv G_*(r,\theta)+g_{*}(\theta),\quad
*=\{r,\theta\},\quad \natural=\{(em),(f)\}
\eea
where $g_{\theta}(r)$ and $g_{r}(\theta)$ are functions to be fixed by the integration.
{Within particular conditions on the fields }
the general integral is
\bea\label{W_q_neq0}
\int\frac{dp}{\rho+p}=-(W^{(f)}+W^{(em)}).
\eea
The Euler equation for this system has been exactly integrated for the background spacetime of Schwarzschild and Kerr \textbf{BHs}
 in  \citep{Komissarov:2006nz,Montero:2007tc} with    a magnetic field is
\bea\label{RSC}&&
B^{\phi }=\sqrt{\frac{2 p_B}{g_{\phi \phi }+2 \ell  g_{t\phi}+\ell ^2g_{tt}}}\quad\mbox{or alternatively},
 \\
 &&B^{\phi }=\sqrt{{2 \mathcal{M} \omega^q}} \left(g_{t \phi }g_{t \phi }-g_{{tt}}g_{\phi \phi }\right){}^{(q-2)/2} V_{eff}(\ell)
\eea
where
\(
p_B=\mathcal{M} \left(g_{t \phi }g_{t \phi }-g_{{tt}}g_{\phi \phi }\right){}^{q-1}\omega^q
\) is the magnetic pressure,
$\omega$ is here the fluid enthalpy, $q$  and $\Mie$ are constant.
(For $q\neq1$ it is
\(
G_r(r,\theta)=G_{\theta}(r,\theta)=G(r,\theta),
\) in Eq.\il(\ref{Eq:wgg})).
We assume moreover a barotropic equation of state.
According to our set-up we  introduce a deformed (magnetized) { Paczy{\'n}ski potential function} and the
  Euler  equation  (\ref{due}) becomes:
\bea\label{Eq:Kerr-case}
&&\partial_{a}\tilde{W}=\partial_{a}\left[\ln V_{eff}+ G\right]\, \mbox{where}
\\
 &&(a\neq0):G(r,\theta)=\Sie \left(\mathcal{A} V_{eff}^2\right)^{q-1}=\Sie\left(g_{{t\phi }} g_{t\phi}-g_{tt} g_{\phi \phi}\right)^{q-1};\\ &&
\nonumber\mbox{and}\;\mathcal{A}\equiv\ell ^2 g_{tt}+2 \ell  g_{t\phi}+g_{\phi \phi },\quad \Sie\equiv\frac{q \mathcal{M} \omega ^{q-1}}{q-1}
\eea
$\Sie$  is a magnetic   parameter, $q$  and $\mathcal{M}$ (magnitude) are constant
  \footnote{The ratio  $\Mie/\omega$  gives the comparison between the magnetic contribution to the fluid dynamics, through $ \Mie $, and  the hydrodynamic   contribution   through its specific enthalpy $\omega$.}.
We therefore consider  the equation for the
\(
\tilde{W}=
\)constant.   The toroidal surfaces, obtained from the equipotential surfaces \citep{Boy:1965:PCPS:,epl}, are provided considering  this function
\bea\label{Eq:goood-da}
&&\widetilde{V}_{eff}^2\equiv V_{eff}^2 e^{2 \Sie \left(\mathcal{A} V_{eff}^2\right){}^{q-1}}=
\\
&&
\frac{\left(g_{t \phi} g_{t \phi}-g_{tt} g_{\phi \phi }\right) \exp \left(2 S \left(g_{t \phi} g_{t \phi}-g_{tt} g_{\phi \phi }\right)^{q-1}\right)}{\ell ^2 g_{tt}+2 \ell  g_{t \phi}+g_{\phi \phi }}=K^2.
\eea
Potential $\widetilde{V}_{eff}^2$, for $\Sa=0$ reduces to  the effective potential ${V}_{eff}^2$ for the non-magnetized case
$V_{eff}$, a function of the metric and the angular momentum $\ell$.
%
The torus  shape is determined by the  equipotential surfaces which now are regulated in eq.\il(\ref{W_q_neq0}) by an effective potential deformed, respect to  the HD case ($B=0$), by the magnetic field.
(We note here that, the magnetic pressure  is regarded  as a   perturbation of  the hydrodynamic component, it is assumed that the  Boyer theory remains   valid and applicable in this approximation --discussions on similar assumptions can be found in \citet{Komissarov:2006nz, abrafra}).

\medskip

\textbf{Leading function: tori distribution law}

\medskip
As for the HD case,  we could  find  the \textbf{\textbf{RAD}} angular momentum distribution:
{\footnotesize{
\bea\label{Eq:dilde-f}
\nonumber
&&\widetilde{\ell}^{\mp}\equiv\frac{\Delta \left(a^3+a r \left[4 \Qa (r-M) \Sie \Delta^{\Qa}+3 r-4\right]\mp\sqrt{r^3 \left[\Delta ^2+4 \Qa^2 (r-1)^2 r \Sie^2 \Delta ^{2 \Qa+1}+2 \Qa (r-1)^2 r \Sie \Delta ^{\Qa+1}\right]}\right)}{
a^4-a^2 (r-3) (r-2) r-(r-2) r \left[2 \Qa (r-1) \Sie \Delta ^{\Qa+1}+(r-2)^2 r\right]}
\\&&\label{Eq:polis-ll}
\mbox{where there is }\, \lim_{\mathcal{\Sie}\rightarrow0}\widetilde{\ell}^{\mp}=\lim_{q\rightarrow 1}\widetilde{\ell}^{\mp}=\ell^{\pm}, \quad \Qa\equiv q-1
\\\nonumber
&& \mbox{or alternatively}\\&&
\quad
\label{Eq:Sie-crit}
\mathcal{\Sie}_{crit}\equiv-\frac{\Delta^{-\Qa}}{\Qa}\frac{a^2 (a-\ell)^2+2 r^2 (a-\ell) (a-2 \ell)-4 r (a-\ell)^2-\ell^2 r^3+r^4}{2  r  (r-1)\left[r (a^2-\ell^2)+2 (a-\ell)^2+r^3\right]}
\eea}}
(dimensionless units)--\citep{Fi-Ringed}.
We provided two different  tori distribution laws. In fact, as noted in \citet{Fi-Ringed}  the introduction of  a toroidal magnetic field $B$, makes the study of the momentum distribution within the disk rather complicated.
In \citet{Fi-Ringed} the critical function  $\Sa_{crit}$  has been introduced as alternative  to represent, instead of $\tilde{\ell}$,  the new leading function for the distribution of tori in the \textbf{RAD} having  a toroidal magnetic field component (each torus is on a line  $\Sie=$constant with  $\ell=$constant value) and able to determine  the limits on the value of the magnetic parameter for the tori formation, the emergence of HD instability associated with the cusped configurations,  $\cc_{\times}$ and $O_{\times}$, and  the emergence of  collision  between two tori of a \textbf{RAD} couple.
 This function, derived from
 $\Sie$-\textbf{\textbf{RAD}} parameter, is
capable of setting  the  location of maximum density points   in the disk and the existence and location  of  the instability points.

 The advantage in using this particular distribution,  not directly connected to the specific momentum distribution of the matter in orbiting rings, is that  it highlights the difference between magnetized  corotating  and  counterrotating tori with respect to the central \textbf{BH}  rotation,   which is   not immediate  with the distribution of momentum $\tilde{\ell}$.
The function  $\Sa_{crit}$ clearly enucleates the  magnetic field contribution in the   $\Qa$ term, while interestingly highlights the role of the parameters  $\ell$ versus $a$--\citep{proto-jet}.
(We also note the dependence  from the quantities $(a\pm\ell)$ which directly connects the specific spin of the central attractor in the limits  $a\in[0,M]$ with the momentum of the fluid. In suitable conditions on the distance of the orbiting  fluid from  the central  attractor and  the configurations symmetries, the ratio $a$ versus $\ell\sqrt{\sigma}$ or $\ell$ versus $a/\sqrt{\sigma}$ are considered).


As demonstrated  in \citet{Fi-Ringed}, the  magnetized tori  can be formed in the  \textbf{eRAD} agglomerates  for   sufficiently small  $(q\Sie)$. The constraints described in Sec.\il(\ref{Sec:constr})  are essentially confirmed  for the magnetized case even  when the pressure force  induces a modification of the effective potential governing the fluid.
Profiles of  $\ell$corotating  tori distributions   are similar independently by the corotation or counterrotation of the fluids in the \textbf{eRAD} with the respect to the central Kerr \textbf{SMBH}. Generally the inner torus has maximum values of the  $\Sie$ smaller then  the maximum of $\Sie$ found in the outer tori.
The most interesting results   emerge in the case of  $\ell$countorrotating couples  where it is clear that  the magnetic profiles for a  couple  $C_-<C_+$, constituting, according to notation of Sec.\il(\ref{Sec:constr}) by a couple of inner corotating torus and outer counterrotating torus, where double accretion occurs, are  radically different from the case $C_+<C_-$ (inner counterrotating and outer corotating torus). The analysis shows also the importance of the  coupling between the toroidal  component of the magnetic field and the fluid angular momentum, particularly in the counterrotating case, $\ell<0$,  for values  $q<1$  when excretion can arise. 

\medskip

\textbf{\emph{Consideration on magnetic fields and choice of parameters: extended range of field parameters.
}}

\medskip

 Some values of $\Qa\equiv q-1$ (dimensionless units) in  $\Sa_{crit}$   may also provide for   excretion {disks} implying   configurations with minimum (unstable points) of   pressure far from the stability points in the momentum distribution and far from  the  attractor, and  with extreme conditions on the magnetic term. These situations are   otherwise typical of special scenarios with a resulting "repulsive"  force, due to the presence of  electrically   charged fluids \citep{Kovar11,charge2,charge3,Kovar:2016kqh,Kovar:2014tla,Slany:2013rml,Schroven:2018agz}, a naked singularity background  \citep{adamek,Stuchlik:2013yca,Stuchlik:2010zz,Stuchlik:2011zza} or a particular cosmological scenario \citep{[68],PPT,Sla-Stu:2005:CLAQG:,[70]}, or also emerging  from hints of the deep structure of  some geometry modifications due to quantum effects resulting in the repulsive force observable, on the large scales of these agglomerates, as engines empowering the excretion tori \citep{Stuchlik:2014jua}.
The range of this parameter is clearly divided into three  regions: \emph{\textbf{ (i) }}$ q> 1 $, with a subrange\emph{\textbf{ (ii) }}with extreme $ q = 2 $; \emph{\textbf{ (iii)  }}$ 0 <q <1 $.
\textbf{\emph{(i)}}For $ q> 1 $  we get  closed surfaces in the  limit $\Sie\approx0 $, which is in agreement with expansion around  $ \Sie=0 $, that is, when the contribution of the magnetic pressure to the torus dynamics is properly regarded as a perturbation  with respect to the hydrodynamics solution. This is realized when an appropriate condition $(\Sie\ll1)$  on the field parameters is satisfied.
The general trend of $\Sa_{crit}$ as function of $r$ is as follows: there is a maximum  such that    $\Sa_{crit}\in [0, \Sa_{\max}]$,   that is $\Sa$ is
bounded below by the  non-magnetized case  and above by  a maximum value $\Sa_{\max}$.  (It can be shown that  the maximum value $\Sa_{\max}$ depends linearly on $q$.).  The presence of a maximum  value is  however  not always satisfied, depending  mostly on the relative rotation of the fluids and also from the \textbf{BH} spin.
\emph{\textbf{ (ii)}}The  case $q=2$ is indeed  interesting because the magnetic field loops  wrap around with toroidal topology along the torus surface.
In fact,    while  $q=1$ is a singular value for $\Sie$,  the magnetic parameter $\Sie$ is negative for $q<1$, where excretion  tori are possible.
{(These special values of  $q$ however  require a careful analysis   on the  characteristic  of the   Komissarov field.).}

%

\medskip

\textbf{On the disks verticality and thickness}

\medskip

It should be  noted  that in the analysis of the disks, even  in the misaligned case of \textbf{RAD}  under suitable conditions,  using results known as  von Zeipel theorem,
 for many of the characteristics of the tori (especially in the equatorial  case) it is sufficient to
 concentrate on the radial  direction, neglecting altogether the analysis  on the poloidal  direction. That is, it is possible to fix the tori and \textbf{RAD} agglomerate verticality through the analysis of radial direction (in all the effects determined by the maximum density/pressure point and central  \textbf{BH}).
 It is possible to "neglect" the verticality of the disk  since here the poloidal component of the magnetic field is suppressed  and it  is still possible in many respects to consider this model a one-dimensional problem along  the radial dimension. This is possible even in the presence of proto-jets, the open limiting configurations that indicate the step for possible  jets  emission along the axis of \textbf{BH} and \textbf{eRAD} tori rotation. This  possibility stands for the   \textbf{eRAD}  and \textbf{RAD} on a spherical attractor background as well. However  what determines the true dimensionality of the morphology  and instability problem of the configurations  are  implied by the integrability conditions on the  force balance equation projected on the spatial dimensions  due to the  tensor  $h^{ab}$.
According to the discussion in  Sec.\il(\ref{Subsection:MHD}),  for the HD \textbf{eRAD}  structures of  Sec.\il(\ref{Sec:Stationar}),  the time-like vector $U^a$ used in the decomposition of the Einstein--Euler equation of Sec.\il(\ref{Sec:ideal}) and the determination of projection spatial supspace with 3D-adapted metric,  are parallel or antiparallel for each toroid (according to their relative   $\ell$corotating or $\ell$counterrotation rotation respectively) and therefore there is only one system for all seeds rings. However, each ring can have different conditions from the constitutive equations  or equations--of--state. In other words, different parts of the orbiting configurations, can be  originated not from a single homogeneous distribution of matter   but are characterized by different sets of thermodynamic conditions.
 Furthermore, it is clear that as we introduced the {leading  function} approach  it is not necessary to study the set of effective  potential functions $V_{eff}^{(i)}$, for each toroid $(i)$ of  the \textbf{RAD},  and  the gradients with the necessary boundary conditions,  but the  function of angular momentum  distributions $\ell^{\pm}(r)$ or alternatively $\Sa_{crit}(r)$ function.
(We also note that in fact we are using $\ell(r)$ or $\Sa_{crit}(r)$ projected on the system equatorial plane $\sigma=1$.).

The  magnetic field and disks rotation are in  strongly  constrained.  In general tori  formation and evolution in \textbf{RADs} depend on the  toroidal magnetic fields parameters.  The role of the central \textbf{BH} spin-mass ratio, the magnetic field and the relative fluid rotation  and tori rotation  with respect the central \textbf{BH}, are crucial elements in determining the accretion tori features,   providing ultimately  evidence of a strict correlation between \textbf{SMBH} spin,  fluid rotation  and magnetic fields in \textbf{RADs} formation and evolution.
Within this set-up the magnetic contribution to the torus dynamics  can be compared to  the  kinetic pressure  in terms of the torus squeezing.
Concerning the disk morphology and the issue of torus squeezing, the torus thickness is here defined as the maximum height of the surface i.e. $h=2x_{max}$ with coordinate $y_{max}$,  where $x_{max}$ is the maximum point of the  torus surface (which is axial symmetric, located on an equatorial plane of the central attractor and with a maximum on  $x_{max}$ while $y_{\max}$ is on the equatorial plane. This point must not be confused with the extreme of the HD pressure or the density although it is related to this.
We then define  the ratio
$
R_s\equiv{h}/{\lambda},
$
as the {squeezing function}  for the torus, where $\lambda$ is the maximum diameter of the torus surface. (We remind that these tori are geometrically thick  i.e. thickness is almost 1). The  lower is $R_s$   and the thinner is the torus, conversely the higher is $ R_s$ and thicker is the torus.
  In \citet{Fi-Ringed}    specific classes of  tori are identified  for  restrict  ranges of magnetic field parameter, that can be observed    around  some specific \textbf{SMBHs} identified by their dimensionless spin. The torus thickness  is not  modified in a  quantitatively significant way by the presence of a toroidal magnetic field although it is much affected by the variation of the  angular momentum. More specifically  one has  to compare  the magnetic contribution to the torus dynamics with respect to the kinetic pressure  in terms of  the torus squeezing and, more in general, the plasma confinement  in the disk toroidal surface.
The torus squeezing function varies with the angular momentum $ \ell $, the  parameter $K$ and the parameter $ \Sie $ (the last evaluates the magnetic contribution to the fluid dynamics respect to   the hydrodynamic   contribution   through its specific enthalpy).
It is shown that  given the presence of a magnetic pressure as a perturbation of
the hydrodynamic component, there are not quantitatively significant effects on the thick disk model; however the analysis showed that the squeezing function  has in general a monotonic trend with $(\ell, K, \Sie)$\footnote{For the hydrodynamic case
the squeezing function  $R_s$ for the Schwarzschild background   increases monotonically with $K$, and at fixed $K$ decreases with $\ell^2$.
This therefore means that the toroidal surface is squeezed on the equatorial plane with decreasing ``energy'' $K$ and  increasing $\ell^2$-- \citep{Fi-Ringed}.
However this trend is reversed, that is
$R_s=h/\lambda<1$ reaching a minimum values of  $R_s\approx0.95$,
for increasing values of $K$, the squeezing  decreases, or decreases until it reaches   a minimum and then increases.
In the  magnetic   case,  the influence of the magnetic field has to be evaluated  through the parameter  $\Sie$.
In general, as for  the case $\Sie=0$,  the torus is thicker  as the $K$  parameter increases, and becomes thinner as fluid angular momentum increases. Furthermore $R_s$ increases with $\Sie$.
For a   small region of  values of $\ell$,
 the torus becomes thinner with increasing of $\Sie$ and viceversa becomes  thicker with decreasing $\Sie$.}.

\section{Concluding remarks}\label{Sec:conclu}
We reviewed   some key aspects of GR-MHD application to High Energy Astrophysics.
We  started  by  discussing   the initial data problem for a most general Einstein-Euler-Maxwell system. We addressed   the problem of  the initial data for this system.  Reduction procedures  were  presented and  the system was then set in  quasi linear  hyperbolic form.
This was taken as the starting point for the analysis of the stability problem for self-gravitating systems with some prescribed  symmetries where  perturbations  of the  geometry part  were also considered on   the quasi  linear hyperbolic onset.
We considered particularly the  discussion following  \citet{first,second}.
As particularization of these special systems  we mentioned the case of  the   GR-MHD and self-gravitating  plasma ball.
This in fact closes the first part of this review dedicated to the most fundamental problems of the existence and uniqueness of a Einstein-Euler-Maxwell equations, and the  stability problem. We stressed particularly the methodological  view points and the main assumptions and implications on the matter fields, currently used to model main objects of the Astrophysics scenario.
 In the second part of this review, we explored  the main   GR-HD  counterparts disks    formulation
 exploring the  geometrically thick disk models. These provide good-fitting   constraints of several
GR-MHD accreting systems on a given background.
     Black holes hosted in  Active Galactic Nuclei  show evidences of  various phases of the  accretion.
      Consequently   to these processes
      the angular momentum orientation of the infalling material  during  the various accretion periods  can be very diversified leading to aggregates of  toroidal structures,  made by orbiting  tori which can be even  misaligned.   The prospect opens the  investigation for  more complex models of accretion extended matter around \textbf{SMBHs}  with an intricated inner structure.
We  review the ringed accretion disks (\textbf{RADs}) which are models of  clusters of   tori orbiting a central super-massive black hole, associated to  complex  instability processes including  tori collision  emergence.   These are formed by geometrically thick, pressure supported,  perfect fluid tori, which can  empower a wide range of activities.
We concluded with the  geometrically thick GR-HD disks gravitating around a  Kerr \textbf{SMBH}  under the influence of a toroidal magnetic field discussing   tori topology and stability, and the formations of multi accretion events.
 The \textbf{RAD} models are  "constraining-models",  providing constraints  as initial data for  dynamical models in  GR-MHD  supported tori for  more complex systems. In many GR-MHD analyses the only GR-HD has proved to be a good comparative model
 to  obtain  constraints  for the set of tori.

In conclusion the topics of this review unravel from the more general  and fundamental ones of  the equations for very general systems with their  Cauchy problems, by exploring  the different techniques used to face the well-posedness of the problem. Applications  to systems with prescribed  symmetries, for example locally rotating or spherically symmetric systems followed.
Finally  we wrote the  general equations by specifying different assumptions  for the case of plasma, focusing finally on accretion disks.

This work does not intend to be an exhaustive review or a complete monograph, for some useful references  see for example \citet{abrafra,Raine,Luci,Ab-Ac-Schl14}.
The topic is enormously vast and there  certainly are several important elements that we have neglected in this discussion focusing  on some selected  aspects more deeply than others. 
 We focused on the  constraints  in Sec.\il(\ref{Sec:ideal})--particularly on regards of the problem of existence and uniqueness of the solutions and their stability --Sec.\il(\ref{Sec:trob}). Secondly  we
 debated the problems faced when considering the perturbation of the gravitation field in the self-gravitating systems, the thermodynamic conditions implied by these  in  Sec.\il(\ref{Sec:terms-cons}), the propagation of the constraints (addressed in a brief discussion in Sec.\il(\ref{Sec:trob})). We  dealt  with
 the constraints in the context of the accretion disks and particularly in the constrained models of thick disks where the symmetries imply an annulment of the evolution equations --Sec.\il(\ref{Sec:la-stafdisc-gtr})-(\ref{Sec:constr})--and in the multiple accreting  systems --Sec.\il(\ref{Sec:intro-RAD}).
 We
 tackled  the  issue of how to manage the validity of assumptions on matter fields into  constitutive  and state  equations--Sec.\il(\ref{Sec:trob})-Sec.\il(\ref{Sec:prez})---the implications on fluids thermodynamics, Sec.\il(\ref{Sec:terms-cons}), especially in the context of fixed symmetries in self-gravitating and toroid  symmetry on a fixed background- - Sec.\il(\ref{Sec:influence}).

\bibliographystyle{jpp}

\begin{thebibliography}{99}
  \bibitem[Abramowicz(1971)]{M.A.Abramowicz} Abramowicz, M. A. 1971
 Acta. Astron. {21}, 81.


 \bibitem[Abramowicz(1985)]{otte0}Abramowicz, M. A. 1985
Astronomical Society of Japan  37,  4,  727-734.









%














  \bibitem[Abramowicz\&Fragile(2013)]{abrafra}
Abramowicz,  M.~A.  \&  Fragile, P.~C. 2013
   Living Rev. Relativity, {16}, 1.





\bibitem[\protect\citeauthoryear{Abramowicz et al.}{1978}]{AJS78}
 Abramowicz,  M.A., Jaroszy{\'n}ski, M.\& Sikora, M. 1978
\aap, {{63}}, 221.



%
%
%

\bibitem[Abramowicz\&Straub(2014)]{Ab-Ac-Schl14}
Abramowicz, M. A.\& Straub, O. 2014, Accretion discs, Scholarpedia {9}(8):2408.


  \bibitem[Adamek\&Stuchlik(2013)]{adamek}Adamek, K.\&  Stuchlik, Z., 2013
 Class. Quantum Grav. {30}, 205007.




  \bibitem[Alcubierre(2008)]{Alc08}
Alcubierre, M.
\newblock {\em Introduction to $3+1$ numerical Relativity},
\newblock Oxford University Press, 2008.

  \bibitem[Alho et al. (2017)]{AlhMenVal10}
Alho, A., Mena, F. C., \& {Valiente Kroon,}  J.~A. 2017
\newblock  Adv.Theor.Math.Phys. 21  857--899.







  \bibitem[\protect\citeauthoryear{Alig et al.}{2013}]{Aligetal(2013)}	
 Alig, C., Schartmann, M., Burkert, A.,\& Dolagapj, K.  2013 \apj, {771}, 119.


{
\bibitem[\protect\citeauthoryear{Allen et al. }{2006}]{Allen:2006mh}
 Allen, S. W., Dunn, R. J. H., Fabian, A.C., et al 2006, 
  \mnras, 1, { 372},  21}.

  \bibitem[\protect\citeauthoryear{Aly et al.}{2015}]{Aly:2015vqa}
 Aly,  H., Dehnen, W., Nixon, C.\& King, A. 2015
  \mnras, { 449}, 1, 65.

\bibitem{Anile89}
 Anile, A.M. 1989
\newblock {\em Relativistic fluids and magneto-fluids: With applications in astrophysics and
plasma physics},
  \newblock  Cambridge UniversityPress, Cambridge, U.K.; New York, U.S.A., 1989.


\bibitem[Anton et al.(2006)]{Anton2006} Anton, L., Zanotti, O., Miralles, J. A.,  Marti, J.M. , Ibanez,  J. M., Font,  J. A.  and Pons, J. A., 2006,
 \newblock  Astrophys.\ J.\  {\bf 637}, 296.






\bibitem[Balbus(2011)]{Balbus2011} Balbus, S. A. 2011 Nat. \textbf{470}, 475.

\bibitem[\protect\citeauthoryear{Balbus\&Hawley}{1998}]{BHawley}
 Balbus, S. A.\& Hawley, J. F. 1998
  Rev. Mod. Phys,. {70}, 1.




\bibitem[Bardeen\&Petterson(1975)]
{BP}Bardeen, J. M.,Petterson, J. A. 1975 Astrophys. J. \textbf{195}, L65.
\bibitem{BarMaaTsa07}
Barrow, J.~D., Maartens, R., \& Tsagas, C.~G., 2007
\newblock Phys. Rep. {\bf 449}, 131.


\bibitem[Baumgarte\& Shapiro(2003)]{BauSha03}
 Baumgarte, T.~W.\& Shapiro, S.~L. 2003
\newblock Astrophys. J. {\bf 585}, 921.



\bibitem[Bekenstein\&Oron(1978)]{10}
  Bekenstein, J., Oron, D., 1978 
Phys. Rev. D 18, 1809,1--71819.

\bibitem[Bekenstein\&Oron(1979)]{11}
Bekenstein, J., Oron, D., 1979 
Phys. Rev. D 19, 2827,1--72837.


\bibitem[Betschart\&Clarkson(2004)]{Bets-Clark04}
Betschart, G.,
and  Clarkson, C. A. 2004
Class. Quantum Grav. {\bf 21}, 5587.
\bibitem[Blaes(1987)]{Blaes1987} Blaes,  O. M., 1987, Mon. Not. R. Astron. Soc. \textbf{227}, 975.


\bibitem[Blaes et al.(2006)]{BAF2006}
Blaes, O. M., Arras,   P., \& Fragile, P. C. 2006 
Mon. Not. R. Astron. Soc., {369}, 1235--1252.

\bibitem[\protect\citeauthoryear{Blanchard et al.}{2017}]{Blanchard:2017zfe}
Blanchard,  P.~K.  {\it et al.} 201
 Astrophys.J. 843,  2, 106.

 \bibitem[Boyer(1965)]{Boy:1965:PCPS:}
Boyer, R.~H. 1965
Proc. Camb. Phil. Soc., {61}, 527.



{
\bibitem[\protect\citeauthoryear{Bugli et al.}{2018}]{Bugli}
Bugli, M., Guilet, J., M{\"u}ller, E., Del Zanna, L., Bucciantini, N., Montero, P. J.\ 2018, \mnras, 475, 108}


\bibitem[Burston(2008a)]{Burston:2007ws}
 Burston, R.~B., 2008a
  Class.\ Quant.\ Grav.\  {\bf 25}  075002.

  \bibitem[Burston(2008b)]{Bur-cqg-08} Burston, R. B. 2008b
Class. Quantum Grav. {\bf 25} 075004.
\bibitem[Burston\&Lun(2008)]{Burston:2007wt}
Burston,   R.~B. and Lun, A.~W.~C. 2008
  Class.\ Quant.\ Grav.\  {\bf 25}, 075003.




\bibitem[\protect\citeauthoryear{Carmona-Loaiza et al.}{2015}]{Carmona-Loaiza:2015fqa}
  Carmona-Loaiza, J.~M., Colpi,  M., Dotti, M.\&Valdarnini, R., 2015,
  \mnras,\  { 453},  2,  1608.

\bibitem[Chakrabarti(1990)]{Chakrabarti0}Chakrabarti, S. K. 1990   Mon. Not. R. Astron. Soc., {245}, 747.


\bibitem[Chakrabarti(1991)]{Chakrabarti}Chakrabarti, S. K. 1991   Mon. Not. R. Astron. Soc., {250}, 7.



%
\bibitem[Choquet-Bruhat(1965)]{Cho65}
Choquet-Bruhat, Y. 1965
\newblock C. R. Acad. Sci. Paris {\bf 261}, 354.




\bibitem[Choquet-Bruhat(2008)]{Cho08}
Choquet-Bruhat, Y.
\newblock {\em General Relativity and the Einstein equations},
\newblock Oxford University Press, 2008.

\bibitem[Choquet-Bruhat\& Friedrich(2006)]{ChoFri06}
Choquet-Bruhat, Y.~\& Friedrich, H., 2006,
\newblock Class. Quantum Grav. {\bf 23}, 5941.

\bibitem[Choquet-Bruhat\& York(2002)]{ChoYor02}
Choquet-Bruhat, Y. \& York, J.~W., 2002
\newblock Lect.\ Notes Phys. {\bf 592}, 29.


\bibitem[Ciolfi\&Rezzolla(2013)]{Cio-Re} Ciolfi, R. \&Rezzolla, L. 2013
  Mon. Not. R. Astron. Soc. Let., {435},  1.  



\bibitem[Clarkson(2007)]{Clarkson07}
Clarkson,  C, 2007,
  Phys.\ Rev.\ D {\bf 76}, 104034.


\bibitem[Clarkson\&Barrett(2003)]{Clark-Bar03}
Clarkson, C. A.
and Barrett, R. K. 2003
 Class. Quantum Grav. {\bf 20} 3855.


\bibitem[Cremaschini et al.(2013)]{Cremaschini:2013jia}
  Cremaschini, C., Kovár, J., Slany, P., Stuchlik, Z. \& Karas, V. 2013
  Astrophys.\ J.\ Suppl., {209}, 15.


\bibitem[Cremaschini\&Stuchlik(2013)]{Cre-Stu13}
 Cremaschini, C. \&  Stuchlik, Z. 2013,
Phys. Rev. E, {87}, 043113.




{
\bibitem[\protect\citeauthoryear{Das et al.}{2017}]{Das:2017zkl}
  Das, U., Begelman, M.~C. \& Lesur, G., 2017,
  \mnras,  {473}, 2791.}

\bibitem[\protect\citeauthoryear{DeGraf et al.}{2017}]{DeGraf:2014hna}
  DeGraf, F., Dekel, A., Gabor,  J.\& Bournaud, F. 2017
  \mnras,\  { 466}, 2,  1462.

{\bibitem[\protect\citeauthoryear{Del Zanna et al.}{2007}]{Del-Zanna}
 Del Zanna, L., Zanotti, O., Bucciantini, N., Londrillo, P. 2007 A\&A 473, 11}.


\bibitem[De Villiers\&Hawley(2002)]{DeVilliers}
 De Villiers ,J-P.\&  Hawley,  J. F. 2002
 Astrophys. J., {577}, 866.




\bibitem[Disconzi(2014)]{Disconzi(2014)} Disconzi,  M. M., 2014,
Nonlinearity \textbf{27}, 1915.

\bibitem[\protect\citeauthoryear{Dyda et al.}{2015}]{Dyda:2014pia}
  Dyda, S., Lovelace,  R.~V.~E., Ustyugova, G.~V., Romanova, M.~M., Koldoba, A.~V 2015
  \mnras,\  { 446}, 613.





\bibitem[Ellis\&van Elst (1998)]{EllEls98}
Ellis, G.~F.~R. \& van Elst,  H. 1998
\newblock NATO Adv. Study Inst. Ser. C. Math. Phys. Sci. {\bf 541}, 1.


\bibitem[Etienne et al.(2010)]{EtiLiuSha10}
Etienne, Z.~B.  Liu, Y.~T., \& Shapiro,  S.~L. 2010
\newblock Phys. Rev. D {\bf 82}, 084031.


\bibitem[Fishbone\&Fishbone(1976)]{FisM76}
Fishbone, L. G.,   Moncrief, V. 1976
 { Astrophys. J.}{, {207}}, 962.


\bibitem[Font(2003)]{Fon03}
Font,  J. A. 2003
Living Rev.\ Relat.,  {{6}},  4.




\bibitem[Font(2007)]{Font2008}
 Font,  J.~A., 2007
   Living Rev.\ Rel.\  {\bf 11}, 7 .

\bibitem[Font\&Daigne(2002a)]{Font:2002bi}
  Font, J. A. \&  Daigne, F. 2002
  Mon.\ Not.\ Roy.\ Astron.\ Soc., {334}, 383.


  \bibitem[Font\&Daigne(2002b)]{F-D-02}
 Font, J. A.,  Daigne, F. 2002
Astrophys. J., {581}, L23-L26.


\bibitem[Fragile et al.(2007)]{Fragile:2007dk}
 Fragile, P. C., Blaes, O. M., Anninois, P.,  Salmonson,  J. D. 2007
Astrophys. J., {668}, 417-429.


 \bibitem[\protect\citeauthoryear{Fragile\&Sadowski}{2017}]{Fragile:2017lbx}
  Fragile, P.~C.~\&Sadowski, A., 2017
  \mnras,\  { 467}, 1838.


 \bibitem[Frank et al.(2002)]{Raine}
 Frank, J.,
 King, A.,
 Raine, D. 2002
 \emph{{Accretion Power in Astrophysics}}, (Cambridge University Press, Cambridge 2002).



\bibitem[Friedrichs(1974)]{Fri74}
Friedrichs, K.~O., 1974,
\newblock Comm. Pure Appl. Math. {\bf 28}, 749.


\bibitem[Friedrichs(1991)]{Fri91}
Friedrich, H., 1991,
\newblock J. Diff. geom. {\bf 34}, 275.

\bibitem[Friedrich(1996)]{Fri96}
Friedrich, H., 1996,
\newblock Class. Quantum Grav. {\bf 13}, 1451.

\bibitem[Friedrich(1998)]{Fri98}
Friedrich, H. 1998
\newblock Phys. Rev. D {\bf 57}, 2317.

\bibitem[Friedrich\&Nagy(1999)]{FriNag99}
Friedrich, H. \& Nagy, G., 1999
\newblock Comm. Math. Phys. {\bf 201}, 619.

\bibitem[Friedrich\&Rendall(2000)]{FriRen00}
Friedrich, H. \& Rendall, A.~D. 2000
\newblock Lect. Notes. Phys. {\bf 540}, 127.




\bibitem[Giacommazo\&Rezzolla(2007)]{GiaRez07}
Giacommazo, B. \& Rezzolla, L. 2007
\newblock Class. Quantum Grav. {\bf 24}, S235.
\bibitem[\protect\citeauthoryear{Gilli et al.}{2007}]{Gilli:2006zi}
  Gilli, R., Comastri, A.\&  Hasinger, G. 2007
  \aap,  { 463},  79.


{\bibitem[\protect\citeauthoryear{Gimeno-Soler\&Font}{2017}]{Gimeno-Soler:2017qmt}
 Gimeno-Soler,  S.\& Font, J.~A. 2017
  A\&A, { 607}, A68}.
\bibitem[\protect\citeauthoryear{Grasso\& Rubinstein}{2001}]{Grasso:2000wj}
  Grasso, D.\&~Rubinstein, H.~R. 2001
  Phys.\ Rept.,\  { 348}, 163.




  \bibitem[Gundlach\& Mart\'{\i}n-Garc\'{\i}a(2006)]{GunGar06}
Gundlach, C. \& Mart\'{\i}n-Garc\'{\i}a,  J.~M. 2006
\newblock Class. Quantum Grav. {\bf 23}, S387.

\bibitem[Guo\&Tahvildar-Zadeh(1999)]{Gu-Sha1999}
 Guo, Y. and Tahvildar-Zadeh, A. S., 1999
Contem. Math.,
\textbf{238}, 151-161.
\bibitem[Hamersky\&Karas(2013)]{Hamersky:2013cza}
 Hamersky,  J.~ \& Karas,  V.~ 2013,
  Astron.\ Astrophys., {32},  {555}.







 \bibitem[Hawley(1987)]{Hawley1987}
Hawley, J. F.   1987
   Mon. Not. R. Astron. Soc., {225}, 677.

\bibitem[Hawley(1990)]{Hawley1990}
 Hawley, J. F 1990
   Astrophys. J., {356}, 580.

\bibitem[Hawley(1991)]{Hawley1991}
Hawley,  J. F.   1991
 Astrophys. J., {381}, 496.
  \bibitem[Hawley et al.(1984)]{Hawley1984}
 Hawley, J. F., Smarr, L. L., Wilson,   J. R. 1984
  %
Astrophys. J., {277}, 296.




 \bibitem[Horst(1990)]{Ho60} Horst, E. 1990 
Commun. Math. Phys. \textbf{126}, 613-633.


 \bibitem[Igumenshchev(2000)]{Igumenshchev} Igumenshchev, I. V., Abramowicz, M. A. 2000  Astrophys. J. Suppl.,{130}, 463.
     \bibitem[Jaroszynski et al.(1980)]{Jaroszynski(1980)}
Jaroszynski,  M.,   Abramowicz, M. A., Paczynski, B. 1980, Acta Astronm., {30},  1.





 \bibitem[\protect\citeauthoryear{Karas et al.}{2014}]{Karas:2014rka}
  Karas, V., Kopácek, O., Kunneriath, D.~\&~Hamersky, J. 2014
  Acta Polytech.,\  { 54},  6,  398.


\bibitem[\protect\citeauthoryear{Karas\&Sochora}{2010}]{KS10}
Karas, V.\&Sochora, V. 2010
\apj,\  {725},  2,  1507--1515.
  {
 \bibitem[\protect\citeauthoryear{Kawakatu\&Ohsuga}{2011}]{Kawa}Kawakatu, N., Ohsuga, K. 2011,
\mnras,  417,  4,  2562-2570}.
\bibitem[King\&Nixon(2018)]{King:2018mgw}
  King, A. and Nixon, C. 2018
  Astrophys.\ J.\  {\bf 857}, 1,  L7.


 {\bibitem[\protect\citeauthoryear{Kiuchi et al.}{2011}]{f-x-Kiuchi}   Kiuchi, K., Shibata, M., Montero, P. J. Font, J. A. 2011, Phys. Rev. Letters, 106, 251102}.








%

\bibitem[\protect\citeauthoryear{Komissarov}{2006}]{Komissarov:2006nz}
Komissarov,  S.~S., 2006,
  \mnras,\  { 368},  993.






  \bibitem[Kovar et al.(2011)]{Kovar11} 
Kovar, J, Slany, P., Stuchlik, Z., Karas, V., Cremaschini, C., Miller, J.C. 2011,
  Phys. Rev.,{D} ,{84}, 8,  084002.

\bibitem[Kovar et al.(2014)]{Kovar:2014tla}Kov{\'a}{\v{r}}, J., Slan{\'y}, P., Cremaschini, C., Stuchl{\'\i}k, Z., Karas, V., Trova, A. 2014
  Phys.\ Rev.\ D, {90},  4,  044029.

\bibitem[Kovár et al.(2016)]
{Kovar:2016kqh}
Kovár, J., Slany, P., Cremaschini, C., Stuchlik, Z. , Karas, V. \& Trova, A. 2016
  Phys.\ Rev.\ \textbf{D} {\bf 93},  12,  124055.

\bibitem[Koz{\l}owski et al.(1979)]{Koz-Jar-Abr:1978:ASTRA:}
  Koz{\l}owski, M., Jaroszy{\'n}ski, M.,  Abramowicz, M.~A. 1998
Astron. Astrophys., {63}, 209.



   \bibitem[Laskyand\&Lun(2007)]{Las-Lun07}
Laskyand, P. D.\&  Lun,  A. W. C. 2007
Phys. Rev.  D {\bf 75}, 104010.

\bibitem[\protect\citeauthoryear{Lei et al.}{2009}]{Lei:2008ui}
  Lei, Q., Abramowicz, M.~A., Fragile, P.~C., Horak, J., Machida, M., Straub, O. 2009
  \aap,\  { 498},  471.


  {
\bibitem[\protect\citeauthoryear{Li}{2012}]{Li:2012ts}
 Li., L.~X. 2012
  \mnras,  { 424}, 1461}.
\bibitem[Lichnerowicz(1967)]{Lichnerowicz67}
  Lichnerowicz, A.
  \newblock {\em  Relativistic hydrodynamics and magnetohydrodynamics},
  \newblock Benjamin, New York, 1967.

  \bibitem[\protect\citeauthoryear{Lovelace\&Chou}{1996}]{Lovelace:1996kx}
  Lovelace,  R.~V.~E\&  Chou, T. 1996
  \apj,   {{468}}, L25.





\bibitem[Lubbe\&Kroon(2013)]{LubbeKroon2011kz}
 Lubbe,C. \& Valiente~Kroon, J.~A. 2013
  Annals Phys.\  {\bf 328}, 1.






%
%

\bibitem[Marklund\&Clarkson(2005)]{Marklund:2004qz}
  Marklund, M. and Clarkson, C. 2005
  Mon.\ Not.\ Roy.\ Astron.\ Soc.\  {\bf 358},  892.












\bibitem[\protect\citeauthoryear{Montero et al.}{2007}]{Montero:2007tc}
 Montero, P.~J., Zanotti, O., Font, J.~A.\&Rezzolla, L. 2007
  \mnras,\  { 378},  1101.


\bibitem[Narayan et al.(1998)]{Narayan:1998ft}
Narayan,  R., Mahadevan, R.\& Quataert, E. 1998
  arXiv:astro-ph/9803141.



 \bibitem[\protect\citeauthoryear{Nixon et al.}{2012}]{NixonKing(2012b)}
  Nixon, N. King, A.  Price, D.\& Frank, J. 2012
\apj, {757},  L24.

{
\bibitem[Oka et al.(2017)]{Oka2017} Oka, T., Tsujimoto, S., Iwata, Y., Nomura, M., \& Takekawa, S.\ 2017  Nature Astronomy, 1, 709.
}







\bibitem[Paczy{\'n}ski(1980)]{Pac-Wii}
 Paczy{\'n}ski, B. 1980
\newblock {Acta Astron.}, {30}, 4.

  \bibitem[Paczy{\'n}ski\& Wiita(1980)]{cc}
Paczy{\'n}ski, B. \& Wiita, P. 1980
Astron. Astrophys., {88}, 23.
\bibitem[Page\&Thorne74(1974)]{Page-Thorne74}
Page, Don N. \& Thorne, Kip S. 1974
\apj, {191}, 499-506.


\bibitem[Palenzuela et al.(2010)]{PalGarLehLie10}
Palenzuela, C., Garrett, D., Lehner, L., \& Liebling, S. 2010
\newblock Phys. Rev. D {\bf 82}, 044045.


\bibitem[\protect\citeauthoryear{Parker}{1955}]{Parker:1955zz}
   Parker,  E.~N. 1955
 \apj, {122}, 293.

\bibitem[\protect\citeauthoryear{Parker}{1970}]{Parker:1970xv}
 Parker,  E.~N. 1970
  \apj,  160, 383.


  \bibitem[Porth et al.(2017)]{Luci}
 Porth, O., Olivares, H., Mizuno, Y., et al. 
 2017 
 Comput. Astro.\&Cosm., 4, 1.



  \bibitem[\protect\citeauthoryear{Pugliese\&Kroon}{2012}]{first}
  Pugliese, D.\&Kroon, J.~A.~V. 2012
  Gen.\ Rel.\ Grav.,\  { 44}, 2785.


\bibitem[\protect\citeauthoryear{Pugliese\&Kroon}{2016}]{second}
   Pugliese, D.\&Kroon, J.~A.~V. 2016
  Gen.\ Rel.\ Grav.\  {\bf 48} 6,  74.



\bibitem[Pugliese\&Montani(2013a)]{epl}
  Pugliese, D. \& Montani, G. 2013
  Europhys.\ Lett., 101, 19001.

  \bibitem[\protect\citeauthoryear{Pugliese\&Montani}{2015}]{pugtot}
  Pugliese, D.\&Montani, G. 2015
  \prd, { 91}, 8,  083011.

 \bibitem[\protect\citeauthoryear{Pugliese\&Montani}{2018d}]{Fi-Ringed}
  Pugliese, D.\&Montani, G. 2018
  Mon.\ Not.\ Roy.\ Astron.\ Soc.\  {\bf 476} 4,  4346.

  \bibitem[Pugliese et al.(2013)]{mnras1}
 Pugliese, D., Montani, G. \& Bernardini, M.~G. 2013
  Mon. Not. R. Astron. Soc., 428 (2), 952.


%
%



%


\bibitem[\protect\citeauthoryear{Pugliese\&Stuchlik}{2015}]{ringed}
  Pugliese, D.\&Stuchlik, Z. 2015
  Astrophys.\ J.s,  { 221}, 25.


  \bibitem[\protect\citeauthoryear{Pugliese\&Stuchlik}{2016}]{open}
  Pugliese, D.\&Stuchlik, Z. 2016
  Astrophys.\ J.s,\  { 223}, 2,  27.



\bibitem[\protect\citeauthoryear{Pugliese\&Stuchlik}{2017a}]{dsystem}
  Pugliese, D.\&Stuchlik,  Z. 2017
  Astrophys.\ J.s,\  { 229}, 2  40.



  \bibitem[\protect\citeauthoryear{Pugliese\&Stuchlik}{2018a}]
{Multy}
     Pugliese, D.\& Stuchl\'{\i}k, Z. 2018a
  Class.\ Quant.\ Grav.\  {\bf 35}, 18,  185008.

\bibitem[\protect\citeauthoryear{Pugliese\&Stuchlik}{2018b}]{long}
  Pugliese, D.\&Stuchlik, Z. 2018b
 JHEAp, { 17}, 1.




\bibitem[\protect\citeauthoryear{Pugliese\&Stuchlik}{2018c}]
{proto-jet}
  Pugliese, D.\& Stuchl{\'{\i}}k, Z. 2018c
  Class.\ Quant.\ Grav.\  {\bf 35}, 10,  105005.
\bibitem[\protect\citeauthoryear{Pugliese\&Stuchlik}{2019}]{letter}
  Pugliese, D.\&Stuchlik, Z. 2019  Eur.\ Phys.\ J.\ \textbf{C} {\bf 79} 4,  288.


\bibitem[Pugliese\&Stuchlik(2020a)]{mnras3}
 Pugliese, D. \&Stuchlik, Z. 2020a
MNRAS, 493,  \textbf{3},  4229-4255.

\bibitem[Pugliese\&Stuchlik(2020b)]{mnras3a}
 Pugliese, D. \&Stuchlik, Z. 2020b
submitted
\bibitem[Pugliese\&Stuchlik(2020c)]{proceeding}
  Pugliese, D. and Stuchlik, Z. 2020c
  arXiv:1910.03925 [astro-ph.HE].



\bibitem[Radice et al.(2014)]{Radice:2013xpa}
Radice,   D., Rezzolla, L. and Galeazzi, F. 2014
 Class.\ Quant.\ Grav.\  {\bf 31},075012.


  \bibitem[Rees et al.(1982)]{Rees1982}
Rees,  M. J., Phinney, E. S., Begelman, M. C.,  Biford,   R. D. 1982
 Nature, {295}, 17



%


\bibitem[Renardy(2011)]{Ren11}
Renardy, M., 2011
\newblock J. Math. Fluid Mech. 2011.

\bibitem[Rendall(2008)]{Ren08}
Rendall, A.~D., 2008
\newblock {\em Partial differential equations in General Relativity},
\newblock Oxford University Press, 2008.







\bibitem[Reula(1998)]{Reu98}
Reula, O. 1998
\newblock Living Rev. Rel. {\bf 3}, 1.


\bibitem[Reula(1999)]{Reu99}
Reula, O. 1999
\newblock Phys. Rev. D {\bf 60}, 083507.





  \bibitem[Rezzolla et al.(2003)]{Rez-Zan-Fon:2003:ASTRA:}
 Rezzolla, L., Zanotti, O., Font,  J.~A. 2003
Astron. Astrophys., {412}, 603.
%
%



%




\bibitem[\protect\citeauthoryear{Schee\&Stuchlik}{2009}]{Schee:2008fc}
  Schee, J.\&Stuchlik, Z. 2009
  Gen.\ Rel.\ Grav.,\  { 41},  1795.

\bibitem[\protect\citeauthoryear{Schee\&Stuchlik}{2013}]{Schee:2013bya}
  Schee, J.\&Stuchlik, Z. 2013
  JCAP, { 1304}, 005.



\bibitem[Trova et al.(2018c)]{Schroven:2018agz}
Schroven,   K.,Trova, A., Hackmann, E. and L\"ammerzahl, C. 2018
  Phys.\ Rev.\ \textbf{D} {\bf 98}, 2,  023017.

 \bibitem[Shafee et al.(2008)]{Shafee} Shafee, R., McKinney, J. C, Narayan, R., Tchekhovosky, A., Gammie,  C. F., McClintock, J. E. 2008
Astrophys. J., {687}, L25.



\bibitem[Shakura(1973)]{[S73]}
Shakura, N. I.  1973 Sov. Astronomy, {16}, 756.

 \bibitem[Shakura\&Sunyaev(1973)]{[SS73]}
 Shakura, N.I.\& Sunyaev, R. A. 1073 Astron. Astrophys., {24}, 337.



 \bibitem[Shibata\&Sekiguchi(2005)]{ShiSek05}
 Shibata, M. \& Sekiguchi, Y. 2005
\newblock Phys. Rev. D {\bf 72}, 044014.



\bibitem[\protect\citeauthoryear{Sikora}{1981}]{Sikora(1981)}
Sikora, M. 1981
\mnras,  {196},  257.



\bibitem[Slany et al.(2013)]{Slany:2013rml}
Slany, P., Kovar, J., Stuchlik, Z. \& Karas, V. 2013
  Astrophys.\ J.\ Suppl.\, {205}, 3.


 \bibitem[Slan{\'y}\&Stuchl{\'{\i}}k(2005)]{Sla-Stu:2005:CLAQG:}
Slan{\'y},  P. \&  Stuchl{\'{\i}}k, Z. 2005
\newblock   Class. Quantum Gravity, {22}, 3623.




\bibitem[\protect\citeauthoryear{Sochora et al.}{2011}]{S11etal}
Sochora, V.,  Karas, V.,  Svoboda, J., Dovciak, M. 2011
\mnras,\  {418},  276--283.

\bibitem[Stewart\&Walker(1974)]{proc}
Stewart, J.M.and Walker, M. 1974 Proc. R. Soc. London \textbf{A} \textbf{431},  49.




\bibitem[\protect\citeauthoryear{Stuchlik}{2005}]{[68]}  Stuchlik, Z. 2005 
 Mod. Phys. Lett. {A}, {20}, 561--75.

 \bibitem[\protect\citeauthoryear{Stuchlik et al.}{2011}]{Stuchlik:2011zza}
  Stuchlik, Z., Hledik, S.\&Truparova, K. 2011
  Class.\ Quant.\ Grav.,\  { 28}, 155017.






\bibitem[\protect\citeauthoryear{Stuchlik\&Kovar}{2008}]{[70]}Stuchlik, Z. \& Kovar, J. 2008  
Int. J. Mod. Phys. {D}, {17},  2089-105.



  \bibitem[\protect\citeauthoryear{Stuchlik\& Schee}{2010}]{Stuchlik:2010zz}
 Stuchlik,  Z.~\&Schee, J. 2010
  Class.\ Quant.\ Grav.,\  { 27}, 215017.


\bibitem[Stuchlik\&Schee(2012)]{Stuchlik:2012zza}
  Stuchlik, Z. \&  Schee, J. 2012
  Class.\ Quant.\ Grav., {29}, 065002.


\bibitem[\protect\citeauthoryear{Stuchlik\&Schee}{2013}]{Stuchlik:2013yca}
 Stuchlik,  Z.~\&Schee, J. 2013
  Class.\ Quant.\ Grav.,\  30,  075012.


  \bibitem[Stuchl{\'{\i}}k\& Slan{\'y}(2006)]{astro-ph/0605094}
 Stuchl{\'{\i}}k, Z.,   Slan{\'y}, P. 2006
  AIP Conf.\ Proc.,{ 861}, 770.


 \bibitem[Stuchl{\'{\i}}k et al.(2000)]{Stu-Sla-Hle:2000:ASTRA:}
 Stuchl{\'{\i}}k, Z., Slan{\'y}, P.,Hled{\'{\i}}k,  S. 2000
\newblock {Astron. Astrophys.}, {363}, 425

\bibitem[\protect\citeauthoryear{Stuchlik et al.}{2009}]{PPT}
Stuchlik, Z., Slany, P.,  Kovar, J. 2009 Class. Quantum Grav., {26}, 215013.  


\bibitem[\protect\citeauthoryear{Stuchlik et al.}{2015}]{Stuchlik:2014jua}
  Stuchlik, Z., Pugliese, D., Schee, J.\&Kucáková, H. 2015
  Eur.\ Phys.\ J.\ C, { 75}, 9, 451.



%
%


 \bibitem[Trova et al.(2016)]
{charge2}
  Trova, A., Karas, V., Slany, P. and Kovar, J. 2016
  Astrophys.\ J.\ Suppl.\  {\bf 226}, 1,  12.

\bibitem[Trova et al.(2018b)]
{charge3}
Trova, A., Schroven, K., Hackmann, E., et al. 2018 
  Phys.\ Rev.\ \textbf{D} {\bf 97}, 10, 104019.

\bibitem[Tsagas(2005)]{Tsa05}
 Tsagas, C.~G. 2005
\newblock Class. Quantum Grav. {\bf 22}, 393.




  \bibitem[van Elsty\&Ellis(1996)]{Els-Ellis96}
van Elsty, H.  and Ellis,  G. F. R. 1996
%
Class. Quantum Grav. {\bf13}, 1099.
  \bibitem[van Putten(1991)]{Put91}
van Putten, M.~H. P.~M. 1991
\newblock Comm. Math. Phys. {\bf 141}, 63.
    \bibitem[van Putten(2002)]{Put98}
van Putten, M.~H. P.~M. 2002
\newblock J. Math. Phys. {\bf 43}--6195.

\bibitem[Viana et al.(1997)]{VCxL97}Viana, R. L.,  Clemente, R. A.  and  Lopes, S. R. 1997
 Plasma Phys. Control. Fusion \textbf{39}, 197.
  {
\bibitem[\protect\citeauthoryear{Volonteri}{2007}]{apite2}Volonteri, M. 2007 \apj, 663, L5.
\bibitem[\protect\citeauthoryear{Volonteri}{2010}]{apite3} Volonteri, M. 2010 A\&AR, 18, 279.
 }



\bibitem[Volonteri et al.(2003a)]{Volonteri:2002vz}
  Volonteri, M., Haardt, F. \& Madau, P. 2003
  Astrophys.\ J.,{ 582}  559.




\bibitem[\protect\citeauthoryear{Volonteri et al.}{2007}]{apite1}Volonteri, M., Sikora, M., Lasota, J.-P. 2007 \apj, 667, 704.






\bibitem[\protect\citeauthoryear{Wielgus et al.}{2015}]{Wielgus}
 Wielgus, M., Fragile, P. C., Wang, Z., \& Wilson, J. 2015 \mnras, 447, 359.

 \bibitem[\protect\citeauthoryear{Yoshizawa et al.}{2003}]{Y.I.I2003}
 Yoshizawa, A., Itoh, S. I.,  Itoh, K. 2003
  \emph{Plasma\&Fluid Turbulence: Theory and Modelling},
  CRC Press.

\bibitem[Zanotti\&Pugliese(2014)]{Zanotti:2014haa}
Zanotti, O. \&  Pugliese, D. 2015
 Gen.\ Rel.\ Grav.,  { 47}, 4,  44.

  \bibitem[Zenginoglu(2003)]{Zen03}
Zenginoglu, A. 2003
\newblock {\em Ideal magnetohydrodynamics in curved spacetime},
\newblock Master thesis, University of Vienna, 2003.




\end{thebibliography}

\end{document}